\documentclass[aps,prd,10pt,twocolumn,superscriptaddress,floatfix,nofootinbib,amsmath,amssymb,altaffilletter,preprintnumbers]{revtex4-1}
\usepackage[colorlinks=true,pdfstartview=FitV,breaklinks=true]{hyperref}

\usepackage{bm}
\usepackage{graphicx}
\usepackage{multirow}
\usepackage{soul}
\usepackage{amsmath}
\usepackage{fontawesome}
\usepackage{mathrsfs}
\usepackage{amssymb}
\usepackage{yfonts}
\usepackage{makecell}
\usepackage{color}
\usepackage{xspace}
\usepackage{url}
\usepackage{verbatim}
\usepackage{mathtools}
\usepackage{upgreek}
\usepackage{amstext}
\usepackage{times}
\usepackage{booktabs}
\usepackage{aas_macros}
\usepackage{tabulary}
\usepackage{tabularx}
\usepackage{etoolbox}

\usepackage[version=4]{mhchem}
\usepackage[utf8]{inputenc}
\usepackage[dvipsnames,table]{xcolor}
\newcolumntype{Y}{>{\centering\arraybackslash}X}
\hypersetup{urlcolor=BlueViolet,
	    citecolor=Plum,
	    linkcolor=PineGreen}
\usepackage[official]{eurosym}
\definecolor{lightgray}{rgb}{0.9,0.9,0.9}	    
\definecolor{green}{rgb}{0,0.5,0}
\definecolor{red}{rgb}{1,0,0}
\definecolor{blue}{rgb}{0,0,0.5}

\long\def\symbolfootnote[#1]#2{\begingroup%
\def\thefootnote{\fnsymbol{footnote}}\footnotetext[#1]{#2}\footnotemark[#1]\endgroup}

\newcommand{\dbd}[2]{\ifmmode \frac{\textrm{d}#1}{\textrm{d}#2}\else $\textrm{d}#1/\textrm{d}#2$\fi}
\newcommand{\pbp}[2]{\ifmmode \frac{\partial#1}{\partial#2}\else $\partial#1/\partial#2$\fi}

\newcommand{\ra}[1]{\renewcommand{\arraystretch}{#1}}

\DeclareMathAlphabet{\mathpzc}{OT1}{pzc}{m}{it}
 \newcommand{\eV}{\text{e\kern-0.15ex V}\xspace}

 \newcommand{\TeV}{\text{T\kern-0.1ex \eV}\xspace}

\DeclareMathAlphabet{\mathpzc}{OT1}{pzc}{m}{it}

\newcommand{\be}{\begin{equation}}
\newcommand{\ee}{\end{equation}}
\newcommand{\bea}{\begin{eqnarray}}
\newcommand{\eea}{\end{eqnarray}}

\begin{document}

\title{The spin axes of globular clusters and correlations with gamma-ray emission}

\author{Ciaran A. J. O'Hare}\email{ciaran.ohare@sydney.edu.au}
\affiliation{School of Physics, The University of Sydney and ARC Centre of Excellence for Dark Matter Particle Physics, NSW 2006, Australia}

\author{Alberto Krone-Martins}\email{algol@uci.edu}
\affiliation{Donald Bren School of Information and Computer Sciences, University of California, Irvine, CA 92697, USA}
\affiliation{CENTRA, Faculdade de Ci\^encias, Universidade de Lisboa, Ed. C8, Campo Grande, 1749-016 Lisboa, Portugal}

\author{C\'eline B\oe hm}\email{celine.boehm@sydney.edu.au}
\affiliation{School of Physics, The University of Sydney and ARC Centre of Excellence for Dark Matter Particle Physics, NSW 2006, Australia}

\author{Roland M.~Crocker}
\email{rcrocker@fastmail.fm}
\affiliation{Research School of Astronomy and Astrophysics, Australian National University, Canberra 2611, A.C.T., Australia}

\smallskip
\begin{abstract}
A growing number of Milky Way globular clusters have been identified to possess a noticeable degree of solid-body rotation. For several clusters, the combination of stellar proper motions and radial velocities allows for 3-dimensional spin axes to be extracted. In this paper we consider the orientations of these spin axes, and ask whether they are correlated with any other properties of the clusters---either global properties to do with their orbits and origin, or internal properties related to the cluster composition. We discuss the possibility of alignments between the spin axes of globular clusters, chemodynamical groupings, and their orbital poles. We also point out a previously unidentified negative correlation between the measured gamma-ray emissivities and the inclination of the globular cluster spins with respect to the line of sight. Given that this correlation is not present in other wavelengths, we cannot conclusively attribute it solely to sampling bias. If the correlation holds up to scrutiny with more data, it may be indicative of sources of anisotropic gamma-ray emission in globular clusters. We discuss the plausibility of such an anisotropy arising from a population of dynamically formed millisecond pulsars with some degree of spin-orbit alignment.
\end{abstract}

\maketitle

\section{Introduction}
\begin{figure*}
\begin{center}
\includegraphics[trim = 0mm 0mm 12mm 0mm, clip, height=0.43\textwidth]{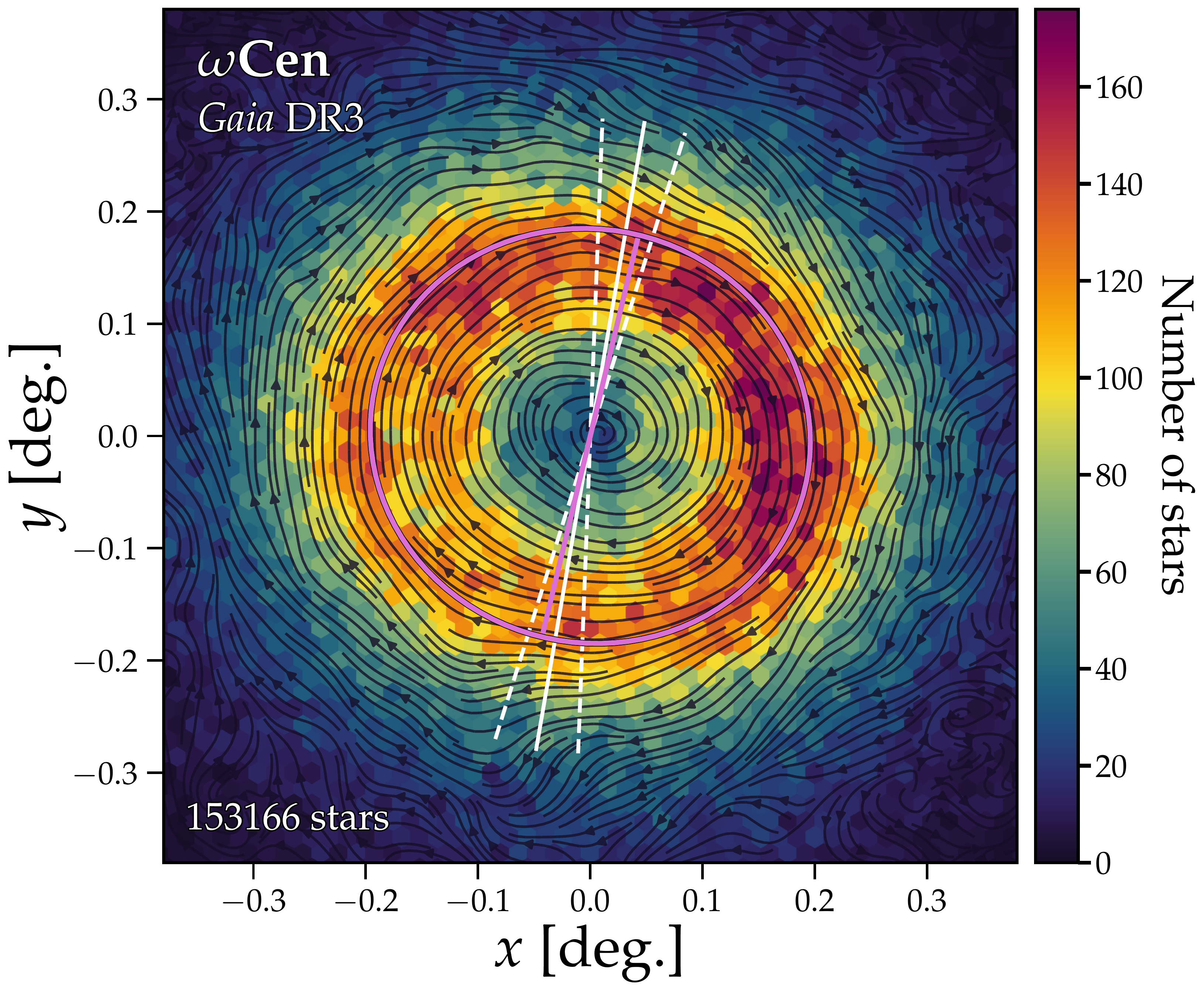}
\includegraphics[trim = 19mm 0mm 0mm 0mm, clip, height=0.43\textwidth]{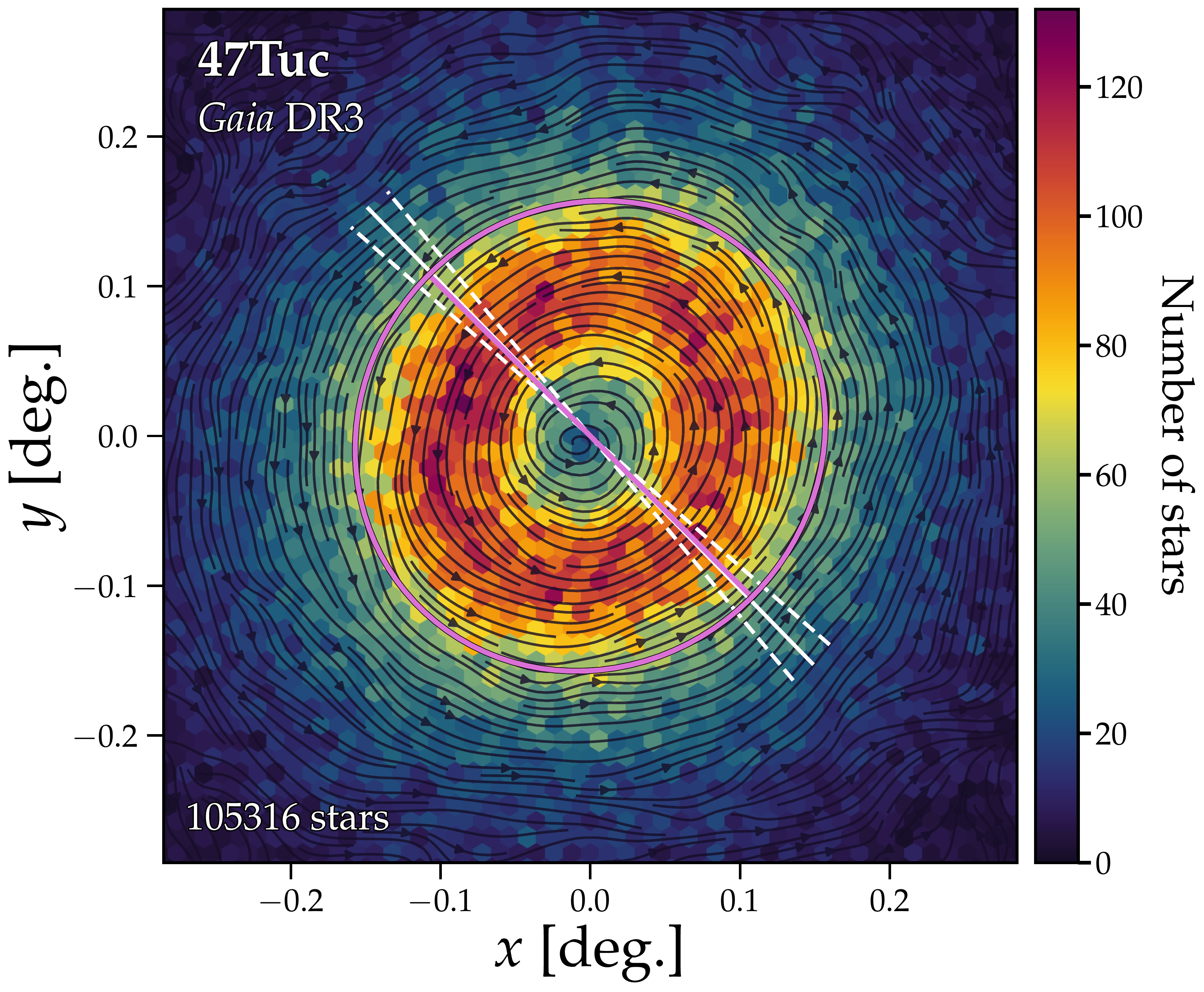}
\caption{Visualisations of the two GCs with the most pronounced rotational signature in the plane of the sky: $\omega$Cen and 47 Tuc. The colour in each case corresponds to the number of stars per hexagonal bin in cluster-centric $(x,y)$ coordinates, which align with right ascension and declination, respectively. The streamlines are derived from the proper motions in the plane of the sky as measured in \textit{Gaia} DR3. The white solid and dashed lines align with the measured position angle of the cluster's rotation axis (and its uncertainty), and the magenta ellipsoid is obtained by fitting the spatial distribution of the stars and has its size chosen to enclose 68\% of the \textit{Gaia} cluster members. We see that the spatial flattening---indicated by the magenta line showing the semi-minor axis---correlates with the kinematic position angle in each of these cases as expected for roughly-spherical rotating systems. Six further clusters are visualised in the Appendix, Fig.~\ref{fig:Streamline_cont}.} 
\label{fig:Streamline}
\end{center}
\end{figure*} 
Despite being abundant across all galaxies and having been observed for centuries, globular clusters (GCs) still hold many closely guarded secrets~\citep[e.g.,][]{doi:10.1146/annurev-astro-091918-104430, doi:10.1146/annurev-astro-081817-051839}. In recent years however, our picture of the internal and global dynamics of the Milky Way's population of GCs has taken several leaps, thanks in large part to the \textit{Gaia} mission~\cite{2016A&A...595A...1G}. \textit{Gaia} has helped reveal and refine our knowledge of internal kinematics~\cite{Sollima_Rotation,Bianchini_RotatingDR2,Vasiliev_Baumgardt_21_EDR3}, complexities in stellar populations~\cite{Cordoni20_OmegaCen}, connections to tidal debris and streams~\cite{Sollima20_TidalTails,Ibata19_Fimbulthul,Simpson20_Fimbulthul,Malhan22_GlobalDynamicalAtlas}, as well as the distributions of globular cluster orbits around the halo~\cite{Callingham_GCgroups}. With the latest data release from \textit{Gaia}~\cite{2022arXiv220800211G} now listing also radial velocities in 111 Galactic GCs~\cite{GaiaDR3_RVs,Gaia_chemical_cartography}, it is likely that this transformation in our understanding of their physical properties, structures, origins, and evolutionary history will continue in the future.

Although we seem to have an increasingly deep
understanding of the possible origins of GCs, 
there are still several outstanding puzzles regarding the Milky Way's population as a whole, as well as with regard to certain individual clusters~\citep[e.g.,][]{doi:10.1146/annurev-astro-081817-051839}. 
For example, $\omega$ Centauri ($\omega$Cen), the largest and brightest GC in the sky, has significant non-luminous mass content~\cite{Evans22_wCen} as well as an unexpectedly high gamma-ray luminosity. These properties, together with its retrograde orbit, observed tidal debris~\cite{Myeong18_ShardsOmegaCen,Ibata19_Fimbulthul,Simpson20_Fimbulthul}, and multiple stellar populations~\cite{Norris96_OmegaCen,Suntzeff96_OmegaCen,Bedin04_OmegaCen,Milone17_OmegaCenStellarPops,Cordoni20_OmegaCen},  have led to debates about its origin and even its classification as a true GC or the remnant of a dwarf galaxy~\cite{BekkiFreedman03_OmegaCen}. 
Questions also remain about whether or not globular clusters possess and/or form within dark matter halos, and how such halos might affect their evolution~\cite{Reynoso22_darkmatterGC,Vitral21_GCMinihalo}. Many GCs, including $\omega$Cen have also now been connected to sources of very high energy emission~\cite[e.g.,][]{Abdo09_47Tuc_Gammas,Zhang16_GammaRay,Song_Fermi_GCs}. X-ray sources like X-ray binaries have been observed for many years in GCs~\cite{Cheng2018_Xrays}, but it has been shown concretely now that some GCs emit powerfully in gamma-rays as well---the latter very likely originating from a population of dynamically-formed millisecond pulsars~\cite{deMenezes:2018ilq,Wu2022_millisecond_pulsars} (but annihilating dark matter has also been put forward as a possibility~\cite{Brown:2018pwq,Brown:2019whs}). Another cluster, Terzan 5, located close to the Galactic centre, has an anomalously high gamma-ray luminosity~\cite{Song_Fermi_GCs}, the largest number of known pulsars~\cite{Martsen22_Terzan5_pulsars}, and, like $\omega$Cen, has a debated origin~\cite{Ferraro_Terzan5, Taylor22_Terzan5}.

One aspect of Milky-Way GCs that we are now able to study in great detail using \textit{Gaia} is their internal kinematics. In particular, the feature that interests us in this work is their \textit{rotation}. There are now between 20 and 30 globular clusters that have been observed to exhibit unambiguous rotation signals~\cite[see e.g.,][]{GaiaDR2_Clusters_and_Dwarfs, Sollima_Rotation, Bianchini_Rotating_HST, Bianchini_RotatingDR2, Vasiliev_Baumgardt_21_EDR3}. For many of these~\cite{Sollima_Rotation}, the evidence of rotation is strong enough in both the plane of the sky and line-of-sight for a full 3D model of a rotating body to be fit and the GC spin to be oriented with respect to the Galaxy. Although the sample size of this set of rotating GCs remains relatively small, what has not been discussed until now is whether or not these spin axes possess any obvious correlations with other measured properties of the clusters. In the present study, we have asked this question broadly and report the existence or otherwise of connections between the spin axes of Milky Way globular clusters and properties that are both intrinsic (i.e.~innate to the cluster itself) and extrinsic (i.e. related to the cluster's orbit and environment).

To be more specific, the properties that we discuss here are grouped into three: i) internal kinematics and rotation, ii) orbital parameters, and iii) luminosities and emissivities. The most interesting correlation that holds up to checks for sampling bias is a negative correlation between the gamma-ray emissivity (luminosity divided by dynamical mass) and the inclination of the cluster with respect to the line of sight---a correlation that 
does not persist in the optical or X-ray bands. The cause for the 
large spread~\cite[e.g.,][]{Song_Fermi_GCs} in measured gamma-ray luminosities of GCs is presently unknown, so a possible dependence on the viewing angle could potentially help to elucidate this large observed spread.

In what follows, we describe the kinematic, orbital dynamics, and luminosities of GCs in Secs.~\ref{sec:kinematics}--\ref{sec:luminosities} for our main sample of clusters that have observed 3D rotation signals at high significance. We also expand our sample to \textit{tentatively} rotating clusters in Sec.~\ref{sec:further}, making use of recent \textit{Gaia} DR3 radial velocities. In Section~\ref{sec:physicalexplanation} we consider in more detail the implications of a correlation between the gamma-ray luminosity and inclination, and what it would imply about the sources of gamma-rays in GCs. Finally, we summarise the observed connections in Sec.~\ref{sec:discussion}, and discuss the implications of the gamma-ray correlation and possible ways to confirm it.

\section{Internal kinematics}\label{sec:kinematics}

\begin{table*}[t]\centering
\ra{1.3}
\begin{tabularx}{0.9\textwidth}{XX|YYYYYYY}
\hline \hline
\multicolumn{2}{c}{{\bf Name}} & $\phi$ [deg.] & $i$ [deg.] & Distance [kpc] & $M/L_V$ [$M_\odot/L_\odot$] & $L_\gamma$ [$10^{34}$ erg s$^{-1}$] & $L_X$ [$10^{32}$ erg s$^{-1}$] & Group \\
\hline
NGC104 & 47Tuc & 224.3$\pm$ 4.6 & 33.6$\pm$ 1.8 & 4.5$\pm$ 0.03 & 1.96$\pm$ 0.09 & 5.61$\pm$ 0.34 & 58.90$^{+7.65}_{-8.56}$ & Disk \\
NGC1904 & M79 & 188.0$\pm$ 10.0 & 37.0$\pm$ 20.0 & 13.1$\pm$ 0.18 & 1.39$\pm$ 0.12 & 2.32$\pm$ 0.98 & 7.42$^{+2.48}_{-2.01}$ & Helmi \\
NGC2808 & - & 36.1$\pm$ 8.4 & 88.5$\pm$ 10.3 & 10.1$\pm$ 0.11 & 1.51$\pm$ 0.06 & 3.43$\pm$ 1.03 & 28.90$^{+2.53}_{-1.86}$ & GSE \\
NGC4372 & - & 226.0$\pm$ 7.0 & 49.0$\pm$ 2.0 & 5.7$\pm$ 0.21 & 2.08$\pm$ 0.12 & $<$1.62 & -- & Kraken \\
NGC5139 & $\omega$Cen & 170.2$\pm$ 7.6 & 39.2$\pm$ 10.3 & 5.4$\pm$ 0.05 & 2.58$\pm$ 0.10 & 3.46$\pm$ 0.42 & 14.10$^{+1.56}_{-1.09}$ & GSE \\
NGC5904 & M5 & 221.6$\pm$ 6.0 & 42.6$\pm$ 3.2 & 7.5$\pm$ 0.06 & 1.81$\pm$ 0.06 & 1.10$\pm$ 0.47 & 5.06$^{+1.37}_{-0.83}$ & Helmi \\
NGC6205 & M13 & 165.5$\pm$ 14.2 & 85.9$\pm$ 11.6 & 7.4$\pm$ 0.08 & 2.31$\pm$ 0.12 & $<$1.57 & 9.15$^{+0.62}_{-0.68}$ & GSE \\
NGC6266 & M62 & 104.2$\pm$ 46.1 & 15.0$\pm$ 12.8 & 6.4$\pm$ 0.10 & 1.99$\pm$ 0.11 & 9.16$\pm$ 0.86 & 63.90$^{+1.02}_{-4.99}$ & Bulge \\
NGC6273 & M19 & 56.9$\pm$ 13.2 & 41.9$\pm$ 7.1 & 8.3$\pm$ 0.16 & 2.25$\pm$ 0.14 & $<$2.28 & -- & Kraken \\
NGC6397 & - & 8.6$\pm$ 15.6 & 72.8$\pm$ 11.9 & 2.5$\pm$ 0.02 & 1.66$\pm$ 0.07 & 0.09$\pm$ 0.05 & 13.30$^{+0.48}_{-0.47}$ & Kraken \\
NGC6541 & - & 83.2$\pm$ 18.3 & 65.4$\pm$ 13.9 & 7.6$\pm$ 0.10 & 1.75$\pm$ 0.08 & 2.10$\pm$ 0.63 & 22.60$^{+0.07}_{-0.08}$ & Kraken \\
NGC6553 & - & 237.7$\pm$ 38.4 & 75.6$\pm$ 29.5 & 5.3$\pm$ 0.13 & 3.22$\pm$ 0.25 & $<$1.38 & 9.96$^{+1.03}_{-0.63}$ & Kraken \\
NGC6626 & M28 & 28.6$\pm$ 17.7 & 83.5$\pm$ 13.3 & 5.4$\pm$ 0.10 & 2.14$\pm$ 0.18 & $<$1.17 & 58.30$^{+0.58}_{-0.53}$ & Bulge \\
NGC6656 & M22 & 252.8$\pm$ 9.2 & 62.1$\pm$ 6.3 & 3.3$\pm$ 0.04 & 2.05$\pm$ 0.08 & $<$0.35 & 4.70$^{+2.18}_{-0.96}$ & GSE \\
NGC6752 & - & 229.1$\pm$ 41.8 & 55.0$\pm$ 20.0 & 4.1$\pm$ 0.04 & 2.34$\pm$ 0.11 & 0.58$\pm$ 0.12 & 11.10$^{+0.38}_{-0.42}$ & Kraken \\
NGC7078 & M15 & 52.6$\pm$ 28.8 & 15.4$\pm$ 5.4 & 10.7$\pm$ 0.10 & 1.58$\pm$ 0.10 & 2.55$\pm$ 0.78 & -- & GSE \\
NGC7089 & M2 & 346.6$\pm$ 12.1 & 52.9$\pm$ 11.2 & 11.7$\pm$ 0.11 & 1.75$\pm$ 0.07 & $<$4.30 & 14.20$^{+6.37}_{-4.01}$ & GSE \\
Terzan5 & -- & 260.4$\pm$ 48.5 & 26.9$\pm$ 34.6 & 6.6$\pm$ 0.15 & 2.53$\pm$ 0.26 & 38.65$\pm$ 2.51 & 117.00$^{+0.84}_{-1.13}$ & Bulge \\
\hline \hline
\end{tabularx}
\caption{Properties of the globular clusters studied in this work, including both their derived spin axis position angles ($\phi$ measured from North to West), inclination with respect to the line of sight $i$, V-band mass-to-light ratio~\cite{Baumgardt20_Vband}, \textit{Chandra} X-ray~\cite{Cheng2018_Xrays} and \textit{Fermi} gamma-ray luminosities~\cite{Song_Fermi_GCs}, and chemodynamical grouping~\cite{Callingham_GCgroups}. The luminosities displayed as inequalities correspond to the 95\% CL upper limits.\label{tab:GCs}}
\end{table*}

\begin{figure*}
\begin{center}
\includegraphics[trim = 0mm 0mm 0mm 0mm, clip, width=0.85\textwidth]{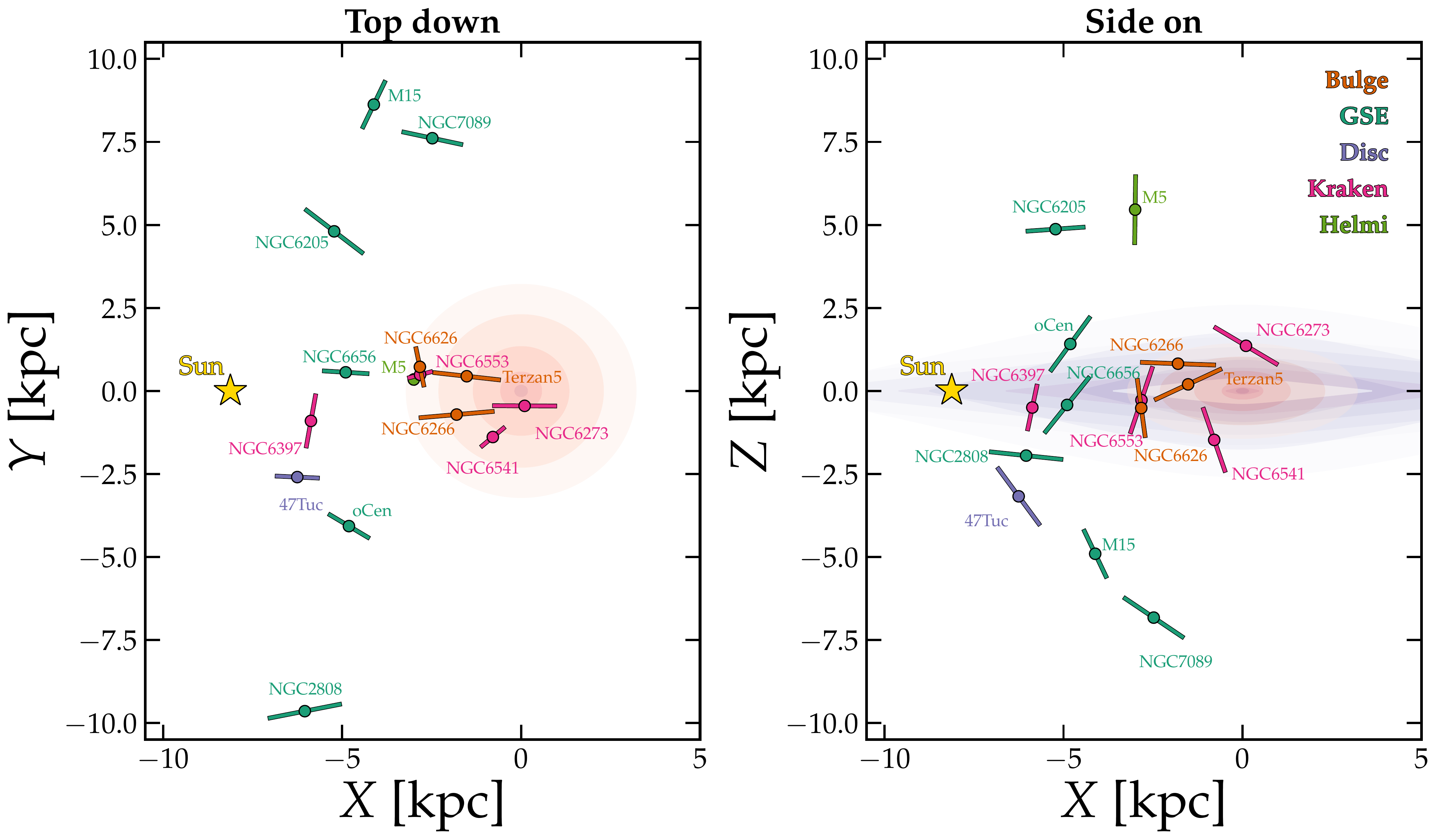}
\caption{Spatial distribution of globular clusters and their measured spin axes in galactocentric coordinates. The left and right panels display a top-down ($X$-$Y$) and side-on ($Z$-$X$) view of the Milky Way respectively. The clusters are coloured according to their names attributed to chemodynamical groupings of stars---some of which align with supposed merger events like the GSE that are believed to have constructed part of the Milky Way halo. The line indicating their spin axis is 2 kpc long in 3D space for all clusters, so the projected length in the plot implies the angle with respect to the page. We also show the log-density profile of the bulge (orange) and thin+thick disk (purple), purely as a visual aid.} 
\label{fig:GC_Galactic}
\end{center}
\end{figure*} 

Globular clusters are typically extremely old ($>10$ Gyr) and present high stellar densities. The numerous star-star interactions implied by both of these properties suggest that GCs should be well-approximated as isotropic, pressure-supported systems with stars arranged along randomised orbits~\cite{King1966}. Indeed this is how internal GC kinematics continue to be broadly understood, although the presence of ordered motion, rotation, anisotropy, and asphericity has been noted and studied for many decades~\cite[see, e.g.,][]{LyndenBell1960_rotatingclusters, KadlaStrugatskaya_M3, McLaughlin_vanderMarel05_clusters}. Understanding these issues has been historically important for extracting accurate measurements of properties such as their masses~\cite{Sollima2015_GCbiases} and non-luminous content, but may also lend insights toward their formation and history.

There are many processes during the formation and lifetime of a GC that can cause it to either gain or lose angular momentum. The hierarchical merger and collapse of clumps within the molecular clouds that initially form GCs will naturally lead to a net rotation at early times~\cite{Mapelli17_rotatingstarclusters}. In principle, this rotation would then be diminished until today as the cluster relaxes and mass is slowly stripped away from it~\cite{Tiongco17_rotation}. On the other hand, this effect must be balanced against the fact that a cluster can also be spun up through interactions with the tidal field of the galactic halo~\cite{Tiongco16_retrogradeclusters, Tiongco_RotatingClusterTidalField}.

As more data has been acquired, a large number of GCs have been shown to display clear evidence for rotation, with amplitudes typically in the few km~s$^{-1}$ range~\cite[see, e.g.,][and references therein]{Pancino07_OmegaCen, Lane11_GCspectroscopy, Fabricius14_CentralRotations, AndersonKing03_47Tuc, Kamann18_MUSEClusters, Bianchini_Rotating_HST, Bianchini_RotatingDR2, Boberg2017_M53, Sollima_Rotation}. This rotation has been observed now in the plane of the sky using astrometric measurements of stellar proper motions obtained by space missions such as~\textit{Gaia}~\cite{2018A&A...616A...1G, 2021A&A...649A...1G, 2022arXiv220800211G} and HST, as well as along the line-of-sight using spectroscopically obtained radial velocities. Both the outer regions of clusters~\cite{Bianchini_RotatingDR2} as well as their centres~\cite{Fabricius14_CentralRotations} show evidence of rotation. In some of the brightest GCs, the signal can be strong enough for complete three-dimensional spin axes to be reconstructed~\cite{Sollima_Rotation}. A number of studies have been conducted over the years to test for correlations between the rotation \textit{amplitudes} and various properties of clusters such as their stellar populations~\cite{Scalco23_RotationvsStellarMass}, or structural properties~\cite{Fabricius14_CentralRotations}. However, so far relatively few studies have addressed the orientations of the spin axes of the population of Milky Way GCs~\cite[e.g.,][]{Piatti_inclinations}.

For visual reference, in Fig.~\ref{fig:Streamline} we display the two GCs which have the strongest rotation signal, which are also the largest and brightest on the sky: $\omega$Cen, 47 Tuc\footnote{Some of the remaining clusters are shown in Fig.~\ref{fig:Streamline_cont}}. We display \textit{Gaia} DR3 stellar positions and proper motions in an $(x,y)$ Cartesian coordinate system that aligns with right ascension and declination respectively. We have transformed the coordinates onto a flat plane using the orthographic projection recommended by the \textit{Gaia} collaboration~\cite{GaiaDR2_Clusters_and_Dwarfs},
\begin{equation}
\begin{aligned}
x=& \cos \delta \sin \left(\alpha-\alpha_{C}\right) \\
y=& \sin \delta \cos \delta_{C}-\cos \delta \sin \delta_{C} \cos \left(\alpha-\alpha_{C}\right) \\
\mu_{x}=& \mu_{\alpha *} \cos \left(\alpha-\alpha_{C}\right)-\mu_{\delta} \sin \delta \sin \left(\alpha-\alpha_{C}\right) \\
\mu_{y}=& \mu_{\alpha *} \sin \delta_{C} \sin \left(\alpha-\alpha_{C}\right) \\
&+\mu_{\delta}\left(\cos \delta \cos \delta_{C}+\sin \delta \sin \delta_{C} \cos \left(\alpha-\alpha_{C}\right)\right)
\end{aligned}
\end{equation}
and have then corrected for perspective contraction by subtracting the expected radial motion induced by the line-of-sight velocity:
\begin{equation}
    \begin{aligned}
&\mu_{x}^{\mathrm{pr}}=-\frac{\pi x}{4.47\times180} \, \bigg(\frac{v_{\rm los}}{{\rm km\, s}^{-1}}\bigg) \bigg(\frac{{\rm kpc}}{D}\bigg) \quad \operatorname{mas~yr}^{-1}\\
&\mu_{y}^{\mathrm{pr}}=-\frac{\pi y}{4.47\times180} \,\bigg(\frac{v_{\rm los}}{{\rm km \,s}^{-1}}\bigg) \bigg(\frac{{\rm kpc}}{D}\bigg) \quad \operatorname{mas~} \mathrm{yr}^{-1}
\end{aligned}
\end{equation}
where $x$, $y$ are the star positions expressed in degrees, $v_{\rm los}$ is the line of sight velocity, and $D$ is the distance to the GC. Perspective correction is important for clusters with large angular size on the sky, and high systemic proper motions, as is the case for e.g.~$\omega$Cen and NGC 3201 respectively.

The colour of the bins in Fig.~\ref{fig:Streamline} indicate the number of stars. The central gap in stellar density is expected for the publicly available \textit{Gaia} data which is incomplete in highly crowded fields like the centres of GCs due to limitations of the number of observed object allocations at the satellite level~\cite{2015A&A...576A..74D}. The streamlines with arrows indicate the direction of the proper motion field, binned on a similar scale to the stellar positions. The cluster members are selected as those with greater than 95\% membership probability according to the membership classification given in~\cite{Vasiliev_Baumgardt_21_EDR3}. In the appendix Fig.~\ref{fig:Streamline_cont} we show similar plots for six further clusters studied in this work that have rotation signatures that are the most apparent by eye. For the remaining set, the rotation signal is present at a statistical level but is less apparent when visualising the cluster on the plane of the sky.

The position angles of the cluster spin axes are traditionally obtained with the asymmetry in the line-of-sight velocities of the stars relative to the systemic motion. However, a useful cross-check of this approach can be simply obtained by considering the orientation of the semi-minor axis of the cluster. We fit the density of stars in the cluster to a rotated ellipsoid of constant ellipticity\footnote{GCs often do exhibit some slight variation in both the position angle of the axes and the ellipticity as a function of radius and also stellar population~\cite[see, e.g.,][]{Milone17_OmegaCenStellarPops}, however for this crude comparison we do not need to consider this, and moreover, for many of the clusters the numbers of stars are too low for this information to be obtained.}. In Fig.~\ref{fig:Streamline} (and Fig.~\ref{fig:Streamline_cont} in the appendix), the result of this fit is shown by a magenta ellipse with its size normalised to enclose 68\% of the cluster members. The straight magenta line indicates the semi-minor axis. The solid and dashed white lines indicate the measured position angle of the cluster's rotation axis and its uncertainty, calculated using line-of-sight velocities (taken from Refs.~\cite{Bianchini_RotatingDR2, Sollima_Rotation, Kamann18_MUSEClusters} and discussed further below). For the majority of our clusters, the semi-minor axis and position angle of the rotation axis align reasonably well within the errors. This is to be expected from the correlation between rotation and flattening predicted in dynamical models of spherical rotating systems~\cite{Wilson75_ellipticaldynamicalmodel} and identified in Milky Way GCs in previous studies~\cite{Fabricius14_CentralRotations,FallFrenk85_flattening,Bianchini13_flattening}.
In Sec.~\ref{sec:further} we will compute the spin axes for several additional clusters that have not been considered previously as they possess only tentative rotation evidence, in those cases this cross-check against the flattening will again be relevant.

A full list of all of the three-dimensional spin axes for the 18 clusters that we study here is given in Table~\ref{tab:GCs}. The most complete sample of these was obtained recently thanks to \textit{Gaia} DR2 data by Sollima et al.~\cite{Sollima_Rotation}. We use two angles $\phi$ and $i$ to describe the 3D spin axis, following the same system as~\cite{Sollima_Rotation}. The angle $\phi$ refers to the position angle of the spin axis on the plane of the sky measured from North to West. It spans the domain 0 to 360$^\circ$ to account for the clockwise/anti-clockwise sense of the rotation, so that a value of $\phi = 0^\circ$ and $\phi = 180^\circ$ both refer to a spin axis projection that aligns parallel to the $y$-direction, but the former refers to clockwise rotation in $x$-$y$ and the latter anti-clockwise. The angle $i$ refers to the inclination of the spin axis with respect to the line of sight (where $i = 0^\circ$ is a cluster rotating totally in the plane of the sky).

This initial list covers only those clusters passing a series of stringent tests for robustness, but there are several more clusters with identified rotation signals (see Table 2 of~\cite{Vasiliev_Baumgardt_21_EDR3}), which failed one or more of the tests. This should be kept in mind when we attempt to draw conclusions based on the population of spin axes, as there will inevitably be biases induced by this selection efficiency. The identification of a 3D rotation signal in a GC relies on the observation of rotation in the plane of the sky \emph{and} the line-of-sight, and both need to be larger than both systematic and random errors. This will inherently bias the resulting sample of spin axes towards brighter clusters and closer clusters, but also towards more massive clusters, which are more likely to have faster rotation speeds.

\begin{figure*}
\begin{center}
\includegraphics[trim = 0mm 0mm 0mm 0mm, clip, width=0.95\textwidth]{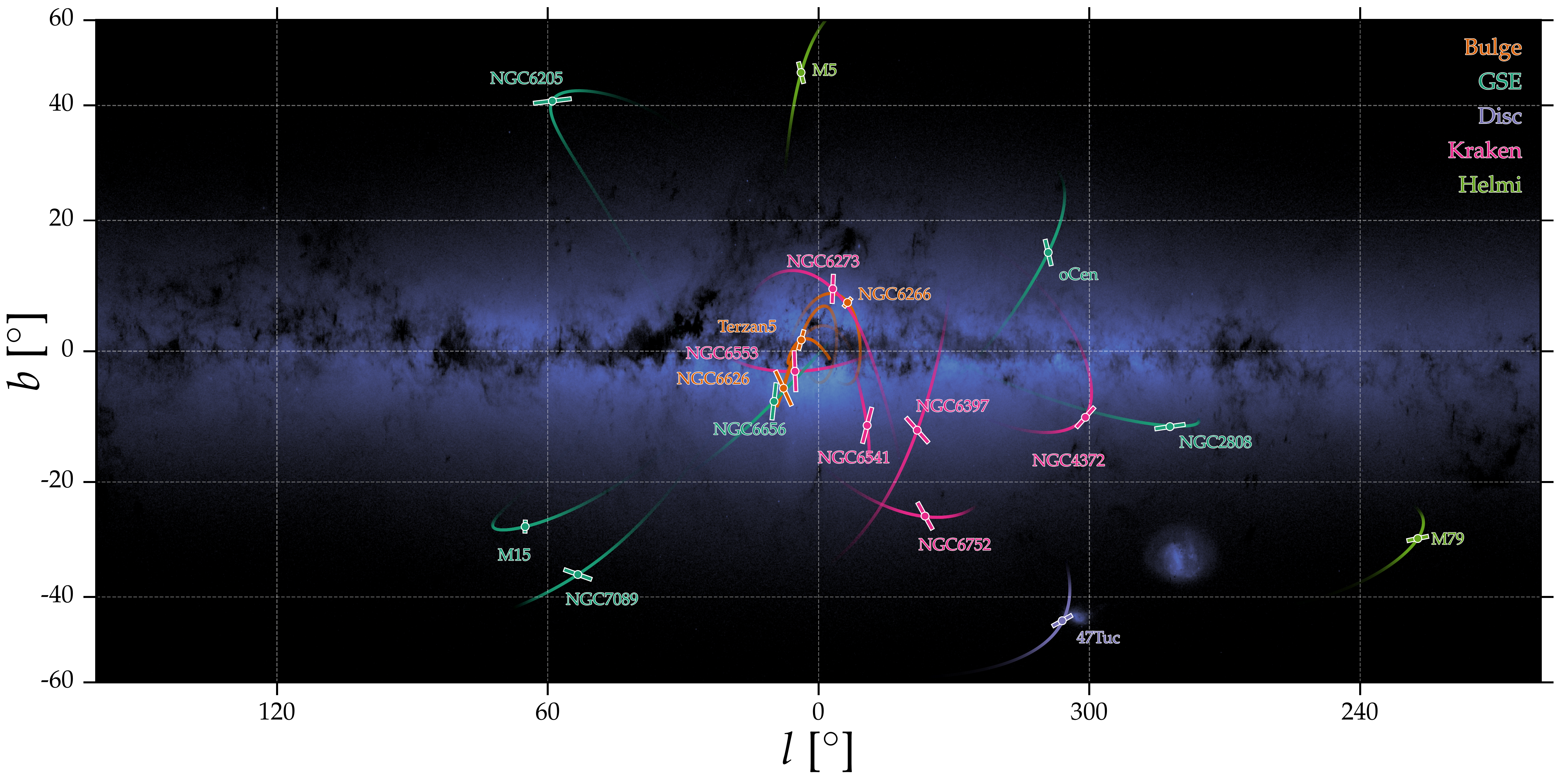}
\caption{A view of the same information as shown in Fig.~\ref{fig:GC_Galactic}, but projected across Galactic longitude and latitude. We also integrate the orbits of the selection of globular clusters forward and backwards in time by a variable amount depending on their orbital speed to display their trajectories.} 
\label{fig:Skymap}
\end{center}
\end{figure*} 

Here we increase our initial sample size slightly by considering three further clusters that only marginally failed robustness checks and have been identified to be rotating in both the plane of the sky and line of sight in previous analyses. The three we include here are NGC 4372, NGC 6752, and NGC 1904. Firstly, the spin-axis position angle of NGC 4372 was measured by~\cite{Kacharov_NGC4372} to be $\phi = 226^\circ \pm 7^\circ$ (converted to the North-through-West system mentioned above), which we can combine with an estimate of the inclination angle of $i = 49$ obtained by~\cite{Bianchini_RotatingDR2}. Using the same method and MUSE line-of-sight observations collected by~\cite{Kamann18_MUSEClusters} we can also assign $\phi = 226.1\pm7$ and $i \simeq 55^\circ$ to the NGC 6752. Finally,~\cite{Leanza22_NGC1904} studied the cluster NGC 1904 (M79) using data from the ESO-VLT MIKiS and obtained $\phi \simeq 188^\circ$ and $i\simeq 37^\circ$. We assign a conservative 20$^\circ$ uncertainty on the inclination angles measured only approximately.
 
This sample of 18 clusters will form the basis of our initial study. In Table~\ref{tab:GCs}, we list all of the measurements mentioned so far in addition to several further properties that will be the subject of the comparisons made in the following sections. These clusters can be claimed with high confidence to be rotating, and their signals are strong enough in all dimensions for a reasonably good reconstruction of a 3D spin axis to be made. To address possible selection biases resulting from the requirements of robustness, in the final section, we will reintroduce clusters with lower significance rotation and extract 3D spin axes with recent \textit{Gaia} DR3 data.

\section{Chemodynamical groups}\label{sec:dynamics}
\begin{figure}
\begin{center}
\includegraphics[trim = 0mm 0mm 0mm 0mm, clip, width=0.49\textwidth]{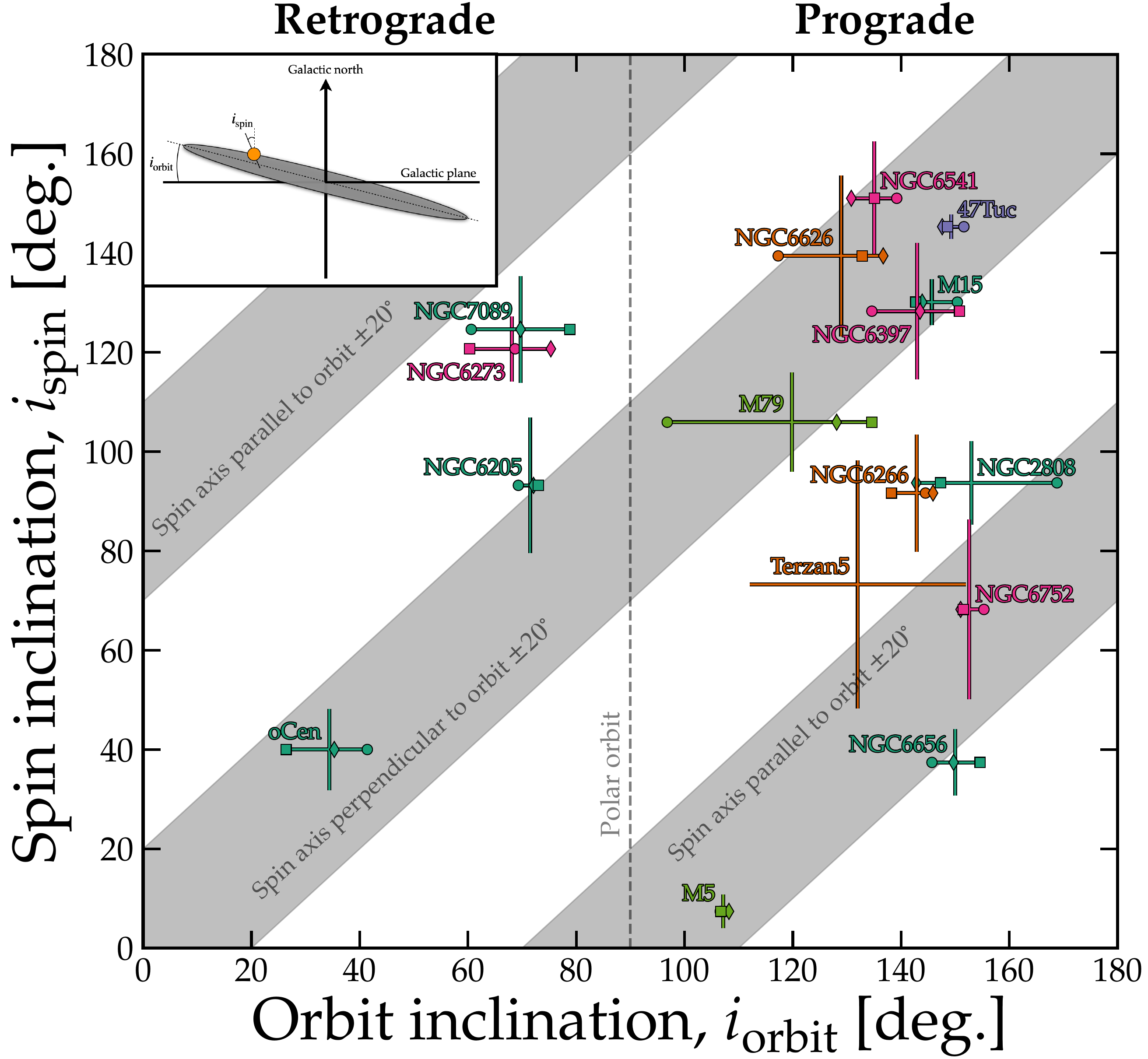}
\caption{Inclination of each GC orbit ($i_{\rm orbit}$) versus the inclination of its spin axis ($i_{\rm spin}$). The inclination of the GC orbit is defined with respect to the Galactic plane, where 90$^\circ$ corresponds to a polar orbit, $<$90$^\circ$ a retrograde orbit, and $>90^\circ$ a prograde orbit---the square, diamond and circle correspond to the three choices for the Milky-Way potential used in~\cite{GaiaDR2_Clusters_and_Dwarfs}. The inclination of the spin axis on the other hand is defined with respect to Galactic north (i.e. a value of 0$^\circ$ corresponds to a GC that aligns with the spin axis of the MW). The diagram in the top left is shown to assist the interpretation of these two angles, though we emphasise that GC clusters do not orbit on simple inclined ellipses, and the inclination angle is really an expression of the fraction of the orbital angular momentum in the galactic $z$-direction. We show grey bands that correspond to cases where the spin axis and the orbital plane are within $\pm 20^\circ$ of being either parallel or perpendicular to each other. The distribution of cluster orbital planes has an excess towards prograde orbits~\cite{GaiaDR2_Clusters_and_Dwarfs}. In addition to this, there is a (very mild) excess of clusters with axes aligned closer to perpendicular to their orbital plane, compared to that expected from randomly aligned spins and orbital planes. We do not confirm the strong evidence of spin-orbit alignment, pointed out in~\cite{Piatti_inclinations}.} 
\label{fig:OrbitInclination}
\end{center}
\end{figure} 

Armed with a set of GC spins, we will now attempt to find or rule out any correlations with other properties and classifications that GCs fall under. 
We begin with phase space groupings. 

Like all stars in the stellar halo, the Milky Way's population of globular clusters can also be grouped according to chemodynamical information. Ever since \textit{Gaia} DR2 generated a huge sample of stars in the nearby stellar halo, numerous studies have been undertaken that ventured to untangle the merger and accretion events that built the Milky Way over its $>10$~Gyr lifetime; see~\cite{Helmi_review} and references therein. In particular, the combination of dynamical information provided by \textit{Gaia}'s astrometry, with chemistry provided by ground-based spectroscopic surveys such as RAVE, APOGEE~\cite{APOGEE_Halo1,Mackereth19_APOGEE_Halo,DiMatteo19_APOGEE_Halo,Horta21_APOGEE_Halo}, GALAH~\cite{GALAH_DR3_StellarHalo}, H3~\cite{Naidu20_HaloH3,Naidu21_GSE} and LAMOST~\cite{Wu22_LAMOST_GSE}, have enabled numerous distinct chemodynamical substructures to be identified and linked to merger events. Each of these events is described by a unique region of phase space described by action-angle coordinates or other integrals of motion. Relevant for this study, globular clusters are also found distributed amongst these groups and, with the use of spectroscopic information, they can further be chemically linked to these merger events, e.g.~Refs.~\cite{Myeong18_GSE_clusters,Callingham_GCgroups}. 

Notwithstanding the above, recent work has cast some serious doubt on how much GCs really align with the chemodynamical phase space groupings that have been found in stellar halo studies~\cite{Pagnini22_GCsNotTracers}, and it is likely that this set of groups are biased by the limited volume over which we can survey the stellar halo~\cite{Sharpe22_MissingStellarHalo}. However, using the naming system as a proxy for chemodynamical information is still a useful tool for us here.

By far the most substantial merger event uncovered by \textit{Gaia} is linked to a substructure now increasingly referred to as the \textit{Gaia}-Sausage-Enceladus (GSE)~\cite{Belokurov_Sausage,Helmi_Enceladus}. The GSE is believed to be a remnant of a dwarf galaxy that merged with the Milky Way around 8--10 Gyr ago~\cite{Naidu21_GSE}. It brought in a large population of stars on highly radially anisotropic orbits that now dominate the inner part of the halo, an event that also influenced the formation of the Milky Way's thick disk~\cite{Belokurov_Splash}. Other distinct substructures have also been discovered and named, and a picture is now emerging of a stellar halo that is almost entirely comprised of accreted material~\cite{Naidu20_HaloH3}, some of which will have GCs associated with it.

GC orbits have been derived previously using \textit{Gaia} data, e.g.,~\cite{Vasiliev19_GC_orbits}, but here we defer to the categorisation of the Milky Way's GCs to that derived by Callingham et al.~\cite{Callingham_GCgroups}. Though in this work they classify GCs into 8 possible groups based on their orbital actions and chemical abundances, the GCs with measured inclination angles that we display here represent only 5 of them: the disk, bulge, GSE, Kraken, Helmi streams~\cite{Helmi99_HelmiStreams,Koppelman_HelmiStreams}, and Sequoia~\cite{Myeong_Sequoia}. The former two are associated with the kinematics of the Milky Way's major components, hence their names, however the latter four are associated with merger events. The GSE is the event that dominates most samples of the inner halo studied so far, however there are also four clusters associated with the Kraken population. The Kraken supposedly stems from another early merger which was suggested to be the origin of a set of old and very low-energy GCs~\cite{Massari19_Kraken,Kruijssen19_Kraken, Kruijssen20_Kraken}, which if real would likely be the Milky Way's earliest major merger event at around 12--13 Gyr ago. Recent studies have suggested that due to the differing levels of impact, the MW tidal field would have on GCs that were accreted at different stages in the construction of the Galaxy, there are correlations between structural parameters and accretion events~\cite{Hammer23_GCaccretion}. 

We can combine the chemodynamical groupings of the GCs and their internal properties, and display them together by visualising their spatial distribution. In Fig.~\ref{fig:GC_Galactic} we display the galactocentric positions and spin axes of the 18 GCs we study here, in both a top-down (X-Y) and side-on (X-Z) perspective. As a visual aid, we also display the logarithmic density profiles of the bulge and thin/thick disk. We colour-code the GCs by the action space grouping. Each plotted spin axis is given the same 3-dimensional length in ($X,Y,Z$), that is, the cluster axes that are pointed into the page appear smaller in length than those that are aligned along the page. In Fig.~\ref{fig:Skymap} we show the same information but projected across the sky in terms of Galactic longitude and latitude. 

The GSE and Kraken are the only populations with enough representation in the sample to be able to extract connections between spin axes. In the GSE case, the clusters are distributed widely across the galaxy and are on both retrograde and prograde orbits and with a large range of inclination angles. On the other hand, the Kraken GCs are distributed closer to the bulge, and spin axes align on average closer to perpendicular to the disk.

In the GSE case, there seems to be some kind of large-scale pattern in the alignment of the spin axes seen in both the X-Y and X-Z projections, potentially indicative of their common trajectory around the Galaxy---this information is related to the phase space position of their orbit which defines the action-space groupings. It is certainly possible that, if the interpretation of the GSE as an accreted dwarf galaxy is correct, then some degree of spin alignment in its globular clusters could be expected. However, a larger sample is needed before any clearer trend can be identified. Currently, the spread of spin axis directions across the galaxy is consistent with an isotropic distribution.

We try to simplify this information slightly while illustrating the distribution in a new way by condensing the cluster's trajectory down to a single number---the orbital inclination. The orbital inclination is defined via the GC's angular momentum projected along the vertical axis of the Milky Way, i.e.
\begin{equation}
\cos{i_{\rm orbit}} = \frac{L_z}{|\mathbf{L}|} \, .
\end{equation}
Here we show the orbit inclination averaged over 10 Gyr when integrated inside three choices for the Milky Way potential used in the \emph{Gaia} DR2 clusters analysis~\cite{GaiaDR2_Clusters_and_Dwarfs}. With the exception of Terzan 5, we neglect observational error in the cluster's present-day systemic motion and position as the uncertainty stemming from the potential by far dominates it---the three symbols correspond to the three choices of potential in those cases. The total orbital angular momentum $\mathbf{L}$ is not conserved along the cluster's orbit in an axisymmetric potential, so we have also checked the instantaneous present-day values of $\cos{i_{\rm orbit}}$, but find broad agreement for the majority of clusters with the integrated value, at least when compared to the uncertainty driven by the choice of potential

In Fig.~\ref{fig:OrbitInclination} we plot the spin axis orientations with respect to Galactic North, against the inclination of the orbit with respect to the Galactic plane, i.e. $i_{\rm orbit}$. We interpret a GC with $i_{\rm orbit} \approx 90^\circ$ as being polar, one with $i_{\rm orbit}<90^\circ$ as being retrograde relative to the disk, and one with  $i_{\rm orbit}>90^\circ$ as prograde. The three symbols correspond to orbits calculated with the three different potentials. As already pointed out in~\cite{GaiaDR2_Clusters_and_Dwarfs}, the clusters are not uniformly distributed in $i_{\rm orbit}$, with a larger number of GCs observed rotating with the disk rather than against.

We see that a large fraction of our sample has spin axes that are aligned consistently with being perpendicular to the planes of their orbits. This could be a consequence of interactions between the GC and the gravitational field of the disk, though it should be kept in mind that GC spin axes are not expected to remain fixed over very long periods. However, for a few clusters, like M5 and NGC6752, their spin axes are actually aligned with the orbital plane, which could also be indicative of some dynamical angular momentum exchange from the disk.

Note that in contrast with some previous suggestions, we do not find any obvious overall correlation between the orientation of the GC spins and their orbital angular momentum. Reference~\cite{Piatti_inclinations} presented evidence of this in terms of the same quantities we have calculated here. However, the apparent negative correlation shown there between two variables defined as $i_{\rm orbit}$ and $i_{\rm spin}-i_{\rm orbit}$ would be present even for isotropically distributed spins and orbital planes.

\section{Apparent Luminosities}
\label{sec:luminosities}
\begin{figure*}
\begin{center}
\includegraphics[trim = 0mm 0mm 0mm 0mm, clip, width=0.49\textwidth]{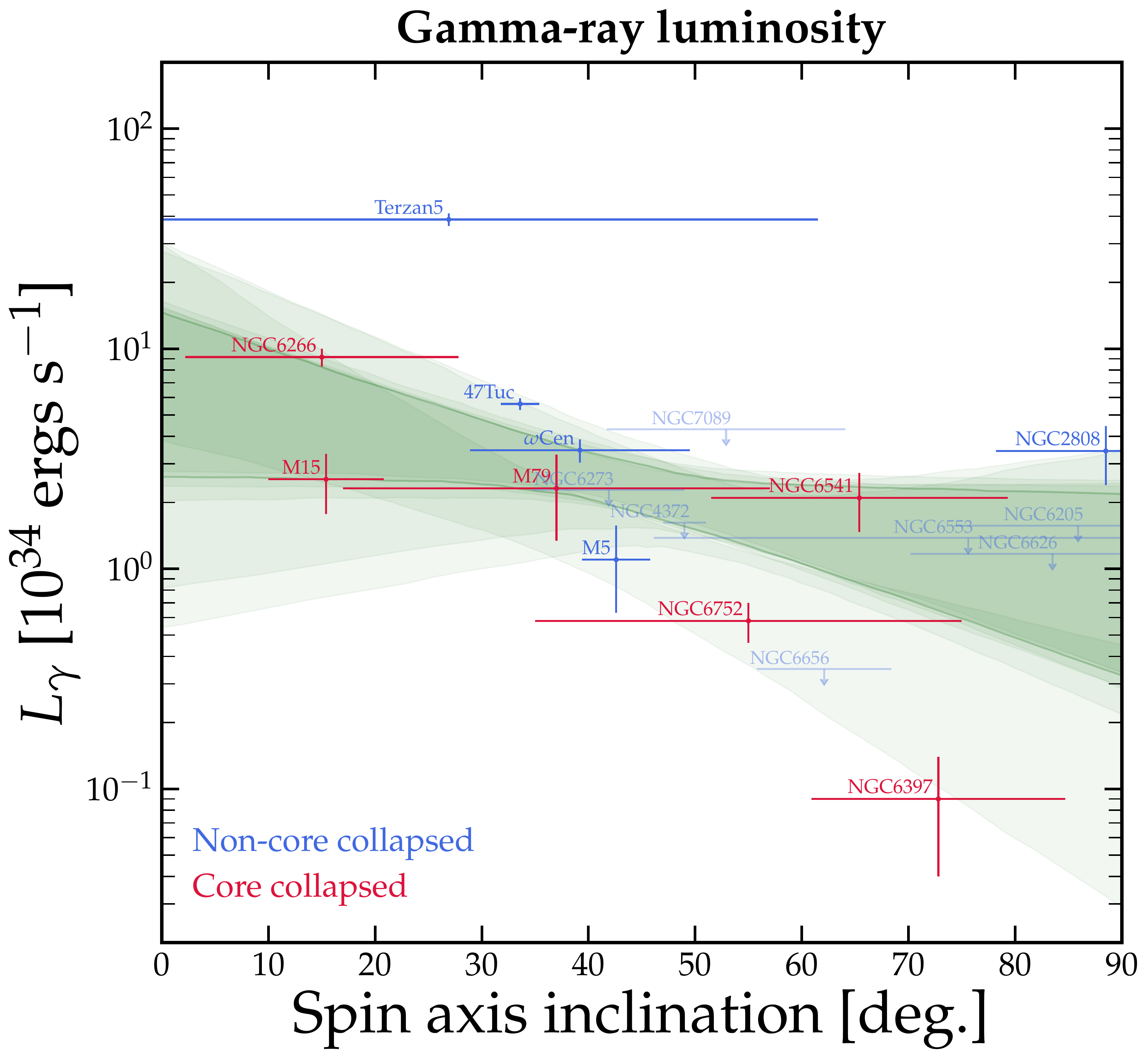}
\includegraphics[trim = 0mm 0mm 0mm 0mm, clip, width=0.49\textwidth]{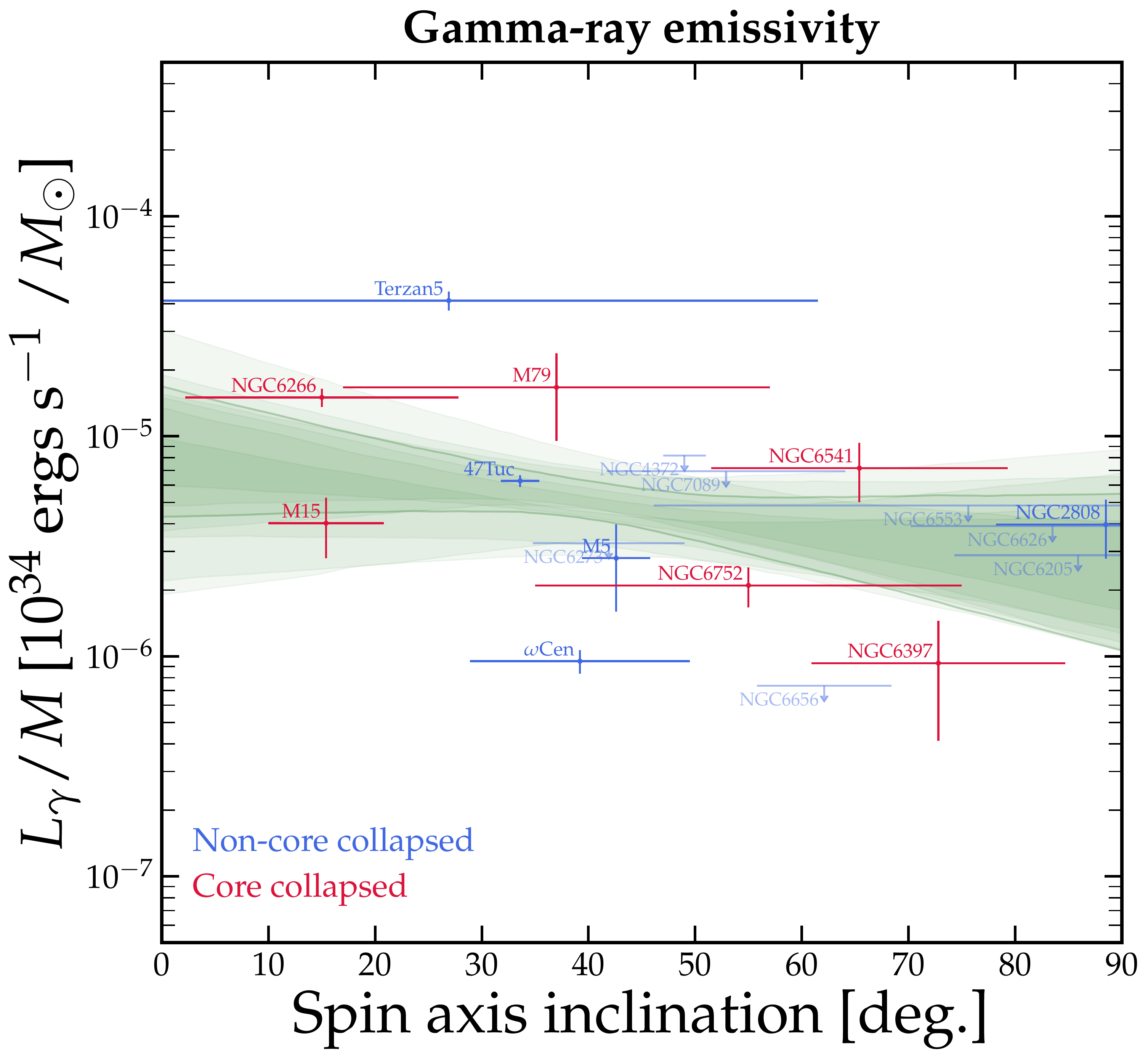}
\caption{{\bf  Left:} The gamma-ray luminosity versus the inclination of the GC spin axis with respect to the line of sight. {\bf Right:} The cluster emissivity (luminosity per dynamical mass) versus inclination. We take the GC gamma luminosities from Refs.~\cite{Song_Fermi_GCs}. In both panels, we show a selection of log-linear fits to the data where the shaded region corresponds to the 95\% CL. The darkest green region is the fit when all data points are used, whereas the other lighter-shaded regions are test cases with two to four clusters removed from the sample at random. In the left panel, the positive gradients are excluded at 97\% CL, whereas in the right panel, positive gradients are excluded at 92 \% CL. Intriguingly this correlation persists even dividing out a possible bias related to the cluster's dynamical mass. 
}
\label{fig:Luminosities_Gamma}
\end{center}
\end{figure*} 

\begin{figure*}
\begin{center}
\includegraphics[trim = 0mm 0mm 0mm 0mm, clip, width=0.42\textwidth]{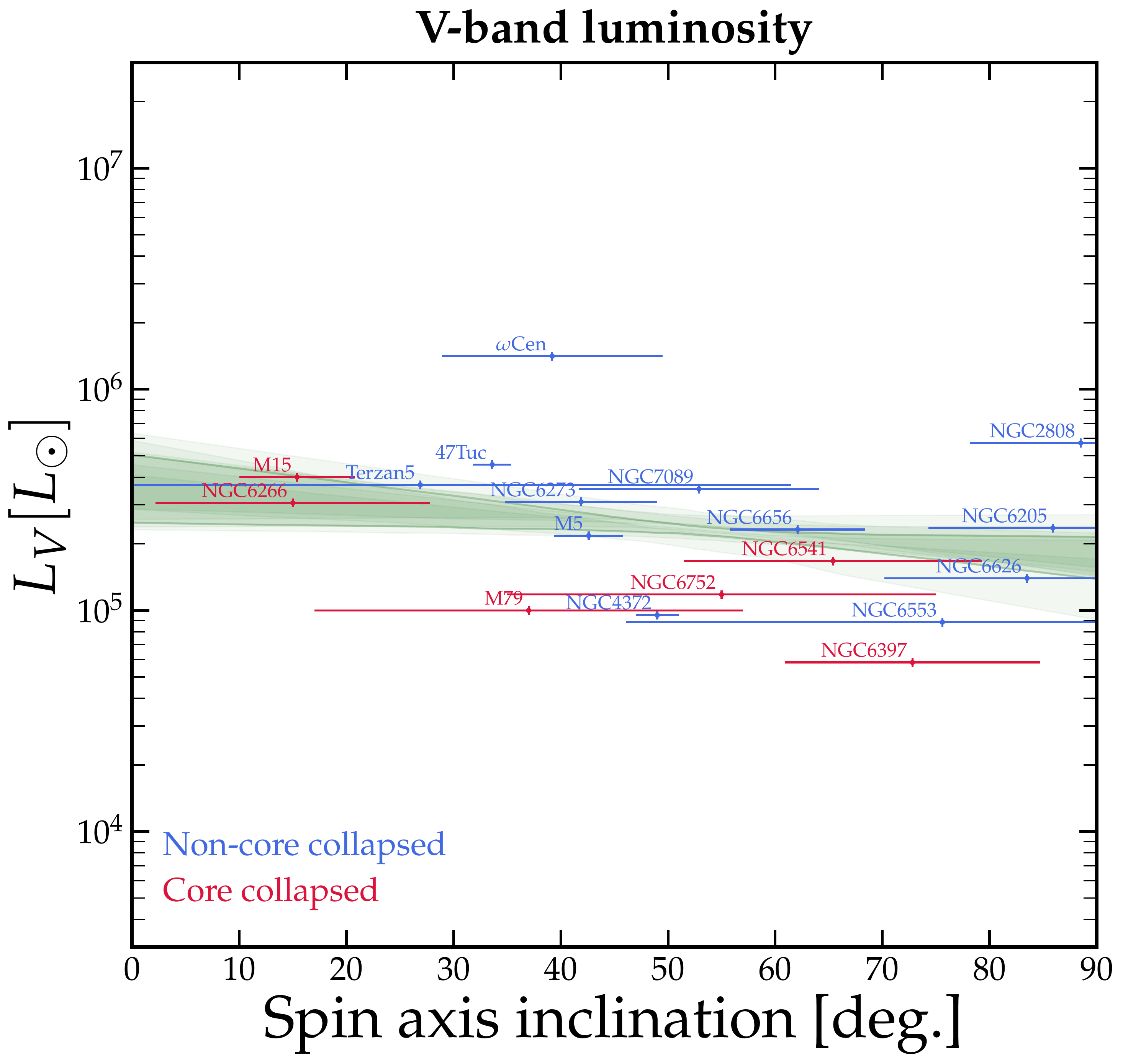}
\includegraphics[trim = 0mm 0mm 0mm 0mm, clip, width=0.42\textwidth]{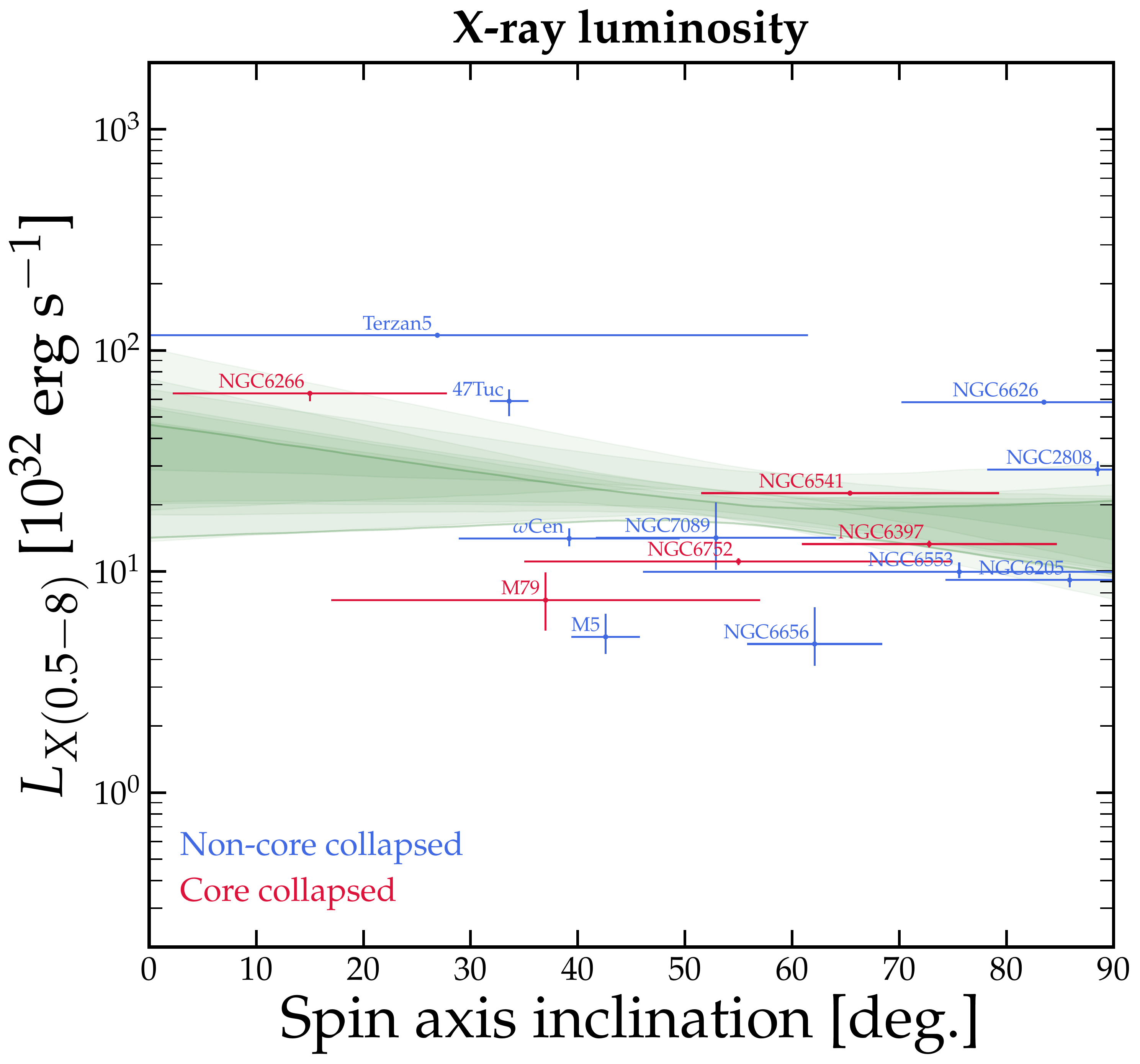}
\includegraphics[trim = 0mm 0mm 0mm 0mm, clip, width=0.42\textwidth]{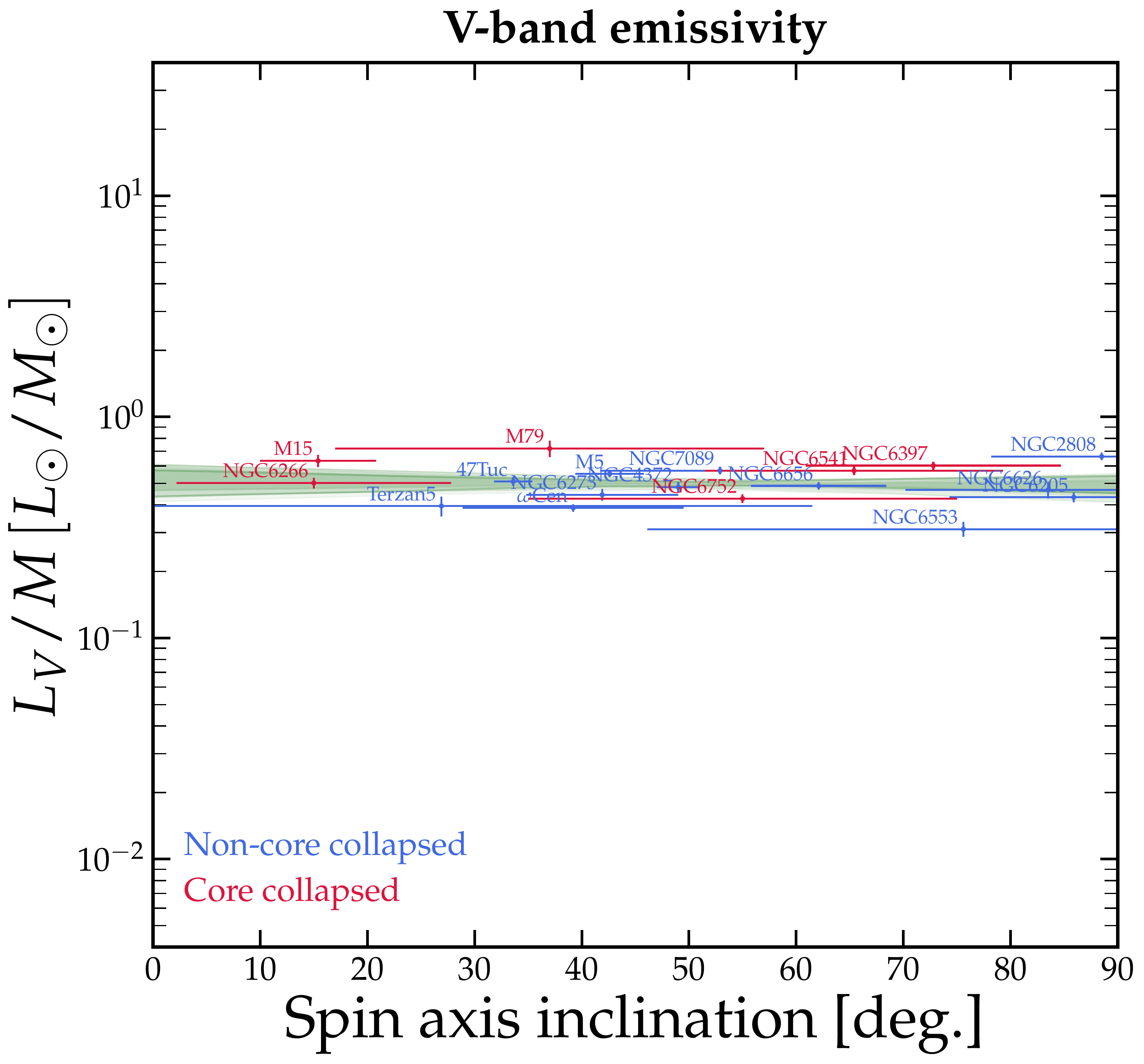}
\includegraphics[trim = 0mm 0mm 0mm 0mm, clip, width=0.42\textwidth]{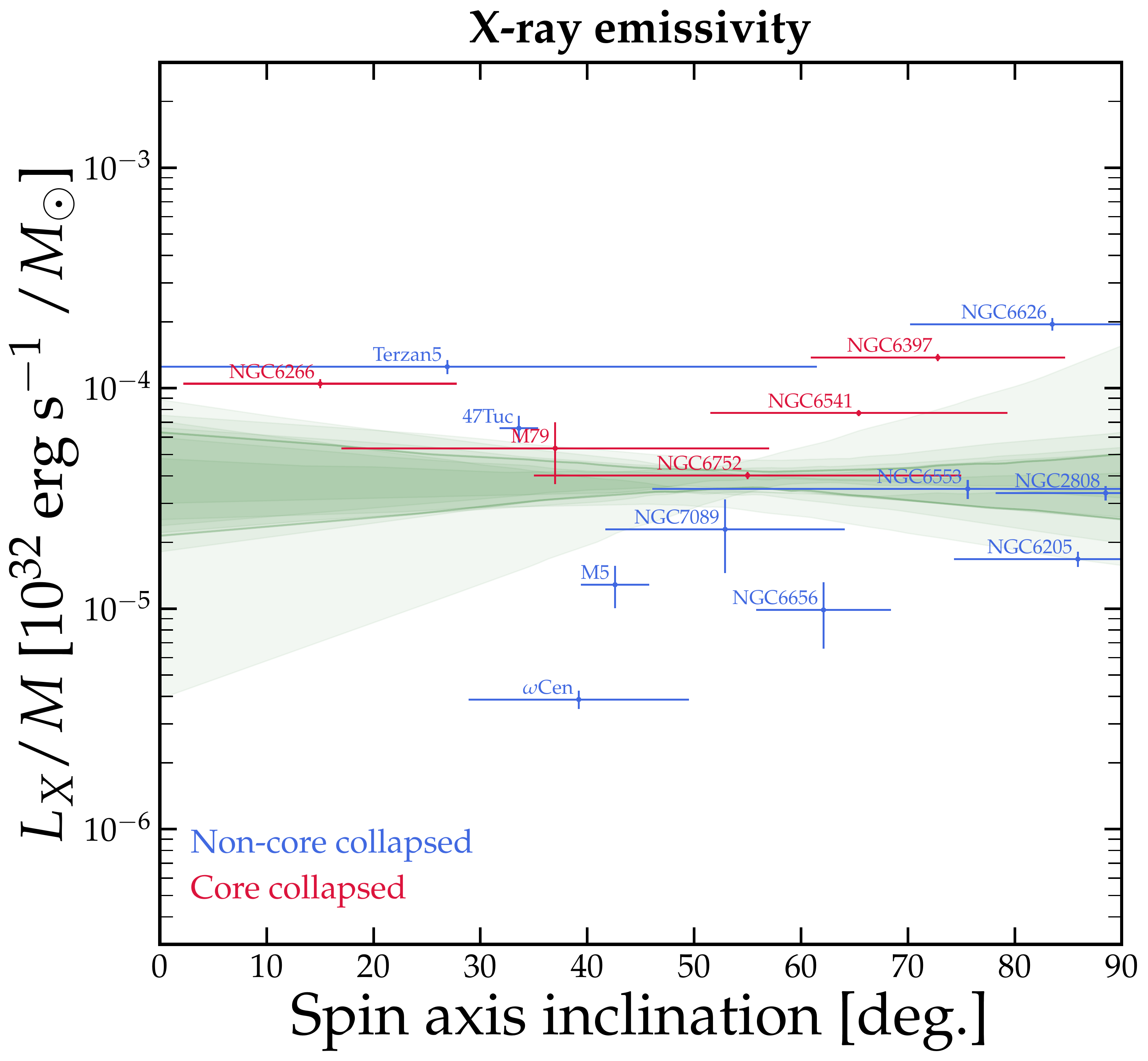}
\caption{{\bf  Top panels:} The luminosities in the V-band (left) and X-ray (right) versus the inclination of the GC spin axis with respect to the line of sight. {\bf Bottom panels:} The same as the top panels but for the cluster emissivity, i.e. luminosity divided by the GC mass. We take V-band and X-ray luminosities from Refs.~\cite{Baumgardt20_Vband,Cheng2018_Xrays} respectively. In each case, we show a selection of log-linear fits to the data where the shaded region corresponds to the 95\% CL. The darkest green region is the fit when all data points are used, whereas the other lighter-shaded regions are test cases with two to four clusters removed from the sample at random. In the upper panels, the positive gradients are excluded at 99 and 89\% CL (from left to right respectively), whereas in the lower panels, positive gradients are excluded at only 72 and 46 \% CL. We have purposefully used the same logarithmic range along the vertical axes in all plots as in Fig.~\ref{fig:Luminosities_Gamma} to further emphasise the large range of luminosities/emissitivies in the case of the gamma-rays compared to the optical and X-ray. We recall that a significant negative correlation persists in the cluster \textit{emissivity} in gamma-rays, but as shown here, it disappears in the emissivities in the V-band and X-ray.}
\label{fig:Luminosities_VX}
\end{center}
\end{figure*} 

Having explored some of the extrinsic characteristics of GCs in the context of their populations around the Galaxy, we now come to an exploration of
how their spins may relate to internal characteristics.

The expectation for rotating GCs generally is that they will spin down over time, with angular momentum transported from the centre of the cluster outwards as the cluster relaxes and loses stars~\cite{Tiongco17_rotation}. Many but not all, clusters were therefore likely rotating faster in the past than in the present day. There has been some discussion of possible trends that the ages and dynamical histories with the GC rotation speed, see e.g.~\cite{Bianchini_RotatingDR2}. More recently, it was observed that more massive stars in GCs have faster peak rotations than less massive stars, as expected in models of multi-mass rotating clusters, see e.g.~\cite{Livernois_RotatingMultimassClusters}.

Here, while surveying possible new correlations between the compositions and spins of clusters, we came across an intriguing relationship between the angle of the GC spin axis with respect to the line of sight and the apparent gamma-ray luminosity inferred from \textit{Fermi} measurements. In this section, we will describe this correlation, show the lack of mirroring correlations in either the optical or X-ray and attempt to determine the reason for the correlation, i.e.~if it is a physical effect or a result of observational bias.

We source gamma-ray luminosities reported by Song et al.~\cite{Song_Fermi_GCs} using the same data set underlying the most recent \textit{Fermi}-LAT 4FGL catalogue~\cite{Fermi_4FGL}. In that work, Song et al.~decisively detected 30 globular clusters in gamma rays. The main source of globular cluster gamma-ray emission is believed to be millisecond pulsars (MSPs). In ~\cite{Song_Fermi_GCs} the authors showed that at a population level---and in some cases, individually---the gamma-ray spectra of globular clusters can be decomposed into two components: an exponentially cut-off power law component peaking at $\sim$few GeV in the spectral energy distribution (SED) and a (generally but not always sub-dominant) pure power-law component. 

The former of these components is generally understood to be magnetospheric
curvature radiation\footnote{Magnetospheric synchrotron radiation has also been proposed as an origin for the \textit{Fermi}-band gamma-ray emission from pulsars; see \cite{Harding2022}; here, for the sake of definiteness, we simply call this magnetospheric component `curvature radiation'.} that is emitted within individual MSP magnetospheres by $\sim$ TeV electrons and positrons ($e^\pm$).

The pure power-law component, on the other hand, was identified as inverse Compton radiation emitted by similarly energetic $e^\pm$ that manage to escape MSP magnetospheres and subsequently radiate off interstellar-medium (ISM) light fields. As we shall return to later, it is notable that GCs are significantly more efficient (per unit stellar mass) than `field' star populations at birthing millisecond pulsars; one potential consequence of this is their significantly enhanced gamma-ray luminosity per stellar mass with respect to these same field populations \cite[e.g.,][]{Song_Fermi_GCs}.

In Fig.~\ref{fig:Luminosities_Gamma}, we plot the gamma-ray luminosities from Song et al.~\cite{Song_Fermi_GCs}, against their measured inclination angles~\cite{Sollima_Rotation}. We observe an apparent by-eye negative or anti-correlation that is persistent against re-sampling the set of clusters. We show a set of five fits performed using orthogonal distance regression~\cite{10.2307/2156807, 10.1145/76909.76913}, accounting for errors on dependent and independent variables. The green bands enclose 95\% CL around the best fit for each re-sample. Each re-sample is a subset of our main sample with three clusters removed at random so as to test for the influence of certain clusters in biasing the fit. We find that a positive or zero gradient in $\log{L_\gamma} \propto i$ is excluded at over 90\% CL in all cases. In Fig.~\ref{fig:Luminosities_Gamma}, we also mark the clusters that have measured inclinations but only upper limits on $L_\gamma$. While these are not included in any of our fits, we do note that they seem to be consistent with the overall trend.

We now wish to find an explanation for this correlation. There are two obvious culprits: a selection bias or some form of anisotropic emission. It is also possible a mixture of the two is at play. We will discuss the arguments for and against these hypotheses in turn. 

Beginning with the possibility of a selection bias, we first have to consider the sample of measured spin axes that we have collated from the studies of Refs.~\cite{Sollima_Rotation,Kacharov_NGC4372,Kamann18_MUSEClusters}. In particular, we should consider closely the selection criteria in the list provided by~\cite{Sollima_Rotation}, which were the most conservative in determining whether a cluster is rotating and provide the largest and most robust sample of spin axes. 

Due to the expected systematics present in the current \textit{Gaia} data release~\cite[e.g.,][]{2021A&A...649A...5F,2021A&A...649A...2L}, rotation amplitudes for clusters are not yet trustworthy if they are below the $\sim$0.1 mas/yr level, even with high-statistics~\cite{Vasiliev2019_ClusterSystematics}. This, therefore, imposes a cut-off in the rotation speed below which a positive detection of rotation cannot be determined in the plane of the sky. Picking up a rotational signal along the line of sight is not subject to such stringent constraints, and the detection of this direction is driven primarily by statistics.

\begin{figure*}
\begin{center}
\includegraphics[trim = 0mm 0mm 0mm 0mm, clip, width=0.475\textwidth]{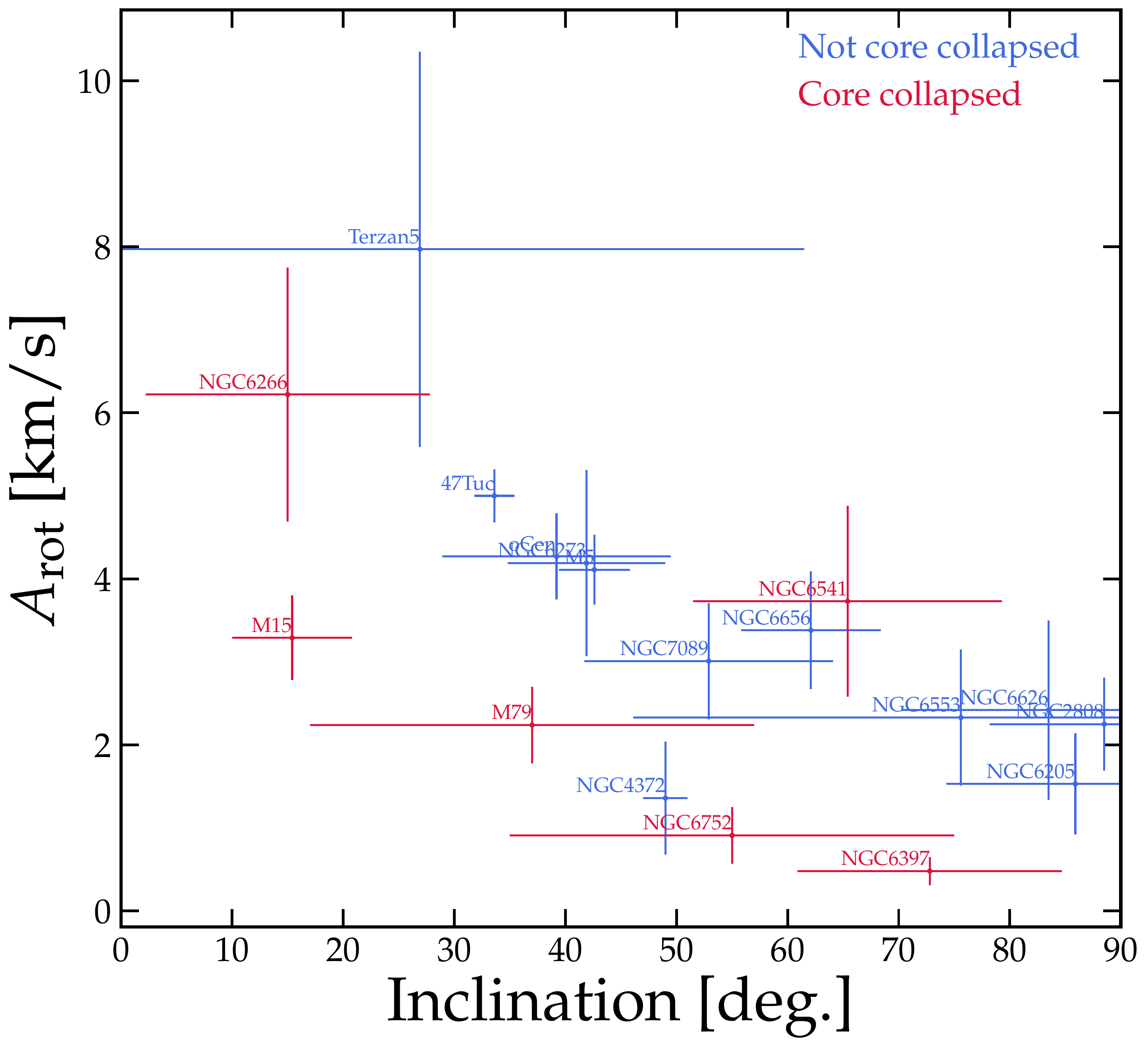}
\includegraphics[trim = 0mm 0mm 0mm 0mm, clip, width=0.49\textwidth]{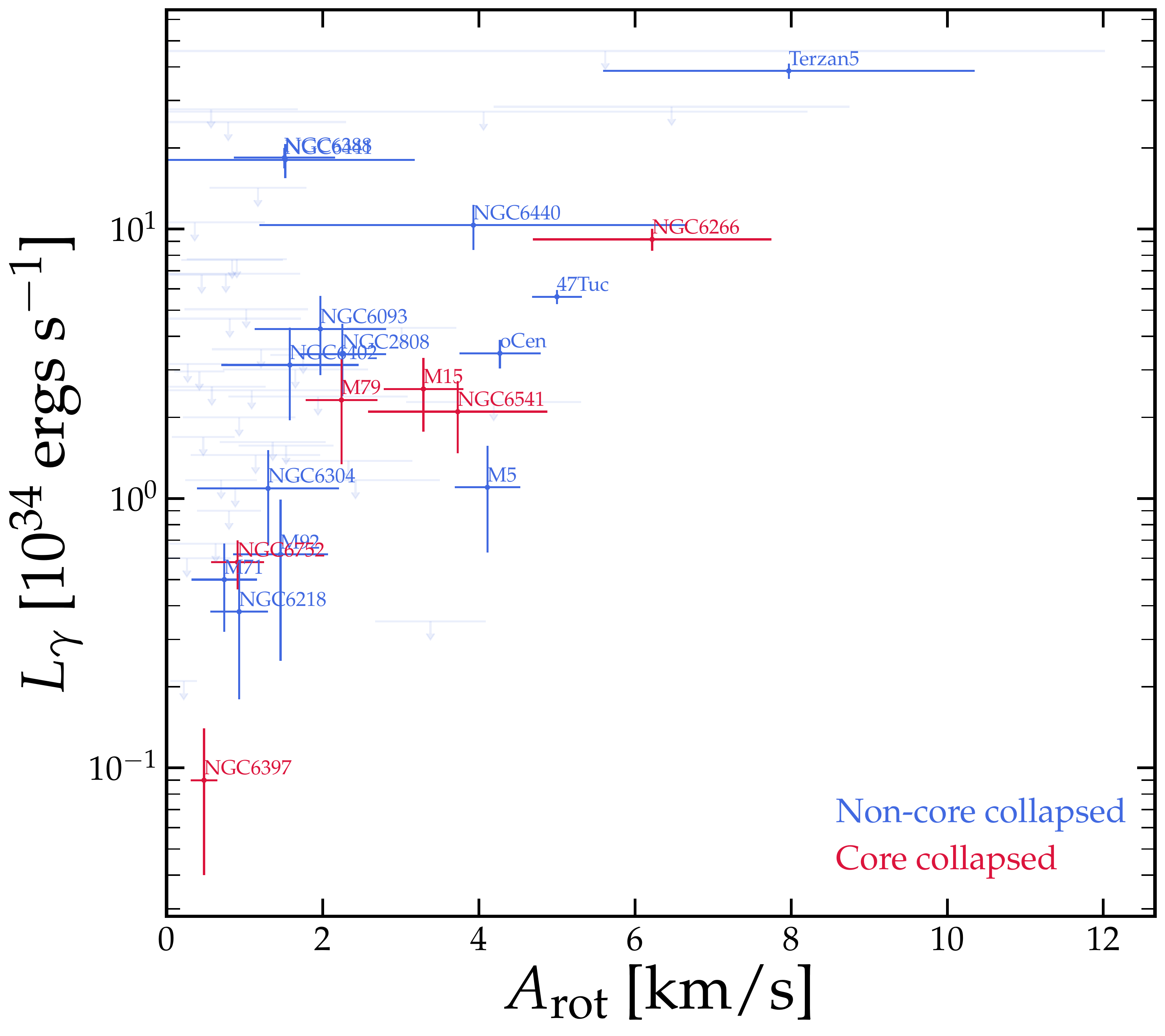}
\caption{{\bf Left}: Amplitude of the GC solid-body rotation speed versus the inclination of the spin axis with respect to the line of sight. {\bf Right}: The rotation amplitude versus the gamma-ray luminosity, with clusters with only upper limits shown as semi-transparent errorbars. The trend shown on the left is purely due to the selection efficiency, where low-inclination low-rotation amplitude clusters are less likely to have robust rotation signals in both the plane of the sky and line-of-sight and hence do not meet the criteria to be included in the sample.} 
\label{fig:AROT}
\end{center}
\end{figure*} 

Therefore, while there will be a bias towards detecting both the plane-of-the-sky and line-of-sight rotations of clusters when they are brighter, only the plane-of-the-sky rotation is further subject to a bias on only detecting \textit{quickly} rotating clusters as well. Therefore we would expect any catalogue of rotating clusters, where a confirmed rotation requires both signals, will generally have a deficit of clusters with $i \lesssim 45^\circ$, due to the additional constraint on seeing rotation with proper motions. Given the rotation speed of clusters correlates with the dynamical mass, and then also loosely with the luminosity, we could expect the detection efficiency to cause us to miss low-luminosity clusters with $i<45^\circ$. This can be seen in the left-hand panel of Fig.~\ref{fig:AROT} where we plot the solid-body rotation speed parameter averaged over the cluster, $A_{\rm rot} = \langle \omega R \rangle$ versus the inclination angle---two variables which should be completely independent. There is clearly an effect present here, stemming from the variable detection efficiency in suppressing the identification of rotation in slowly-rotating clusters with small inclinations. This bias should be expected to feed into the left-hand of Fig.~\ref{fig:Luminosities_Gamma}, however, it remains unclear why the correlation persists reasonably strongly even after we divide by the mass, which also correlates with $A_{\rm rot}$.

In other words, if it were a selection-efficiency bias that is driving the correlation, it would be because we are effectively missing clusters occupying the lower left-hand corners of the plots in Fig.~\ref{fig:Luminosities_Gamma}. Two counterpoints can be made against this however: i) a bias cannot explain the deficit towards the upper-right-hand corner---high-inclination high-luminosity clusters---which should pass the selection criteria more easily than others; ii) this deficit is made difficult to explain statistically because 
the probability distribution of completely random inclination angles should be distributed $\propto \sin{i}$, so missing inclination angles would statistically be weighted towards larger $i$ rather than smaller.

The best approach to further probe this idea would be to increase the considered cluster sample, collecting more clusters, even if they are poorly measured, and verify if the correlation persists or disappears. We will therefore increase the sample to include clusters that have more tentative evidence of rotation in the next section. So, for now, we confront the hypothesis that there is truly some form of intrinsic anisotropy to the radiant intensity of these clusters that causes them to be brighter when viewed along their rotation axes. 

Before jumping to a physical explanation, however, one obvious verification we should make is to analyse if the trend persists at other wavelengths. To remain consistent, we need to take measures of the total GC luminosity instead of the luminosities of point sources inside the clusters. So to do that, we source total V-band luminosities from the catalogue of ~\cite{Baumgardt20_Vband}, as well as the luminosity of weak X-ray sources from \textit{Chandra}~\cite{Cheng2018_Xrays}, which focused on the 0.5--8 keV range for emission from cataclysmic variables and coronally active binaries.

The results in the V-band and X-ray are displayed in the left and right columns of Fig.~\ref{fig:Luminosities_VX}. We note that the logarithmic range of luminosities is much smaller in the optical and X-ray bands, which we highlight in Fig.~\ref{fig:Luminosities_VX}, by using the same logarithmic range in the vertical axis of all four panels as was used in the two panels of Fig.~\ref{fig:Luminosities_Gamma}.

We have highlighted clusters that are classified as core-collapsed (i.e., very high central stellar densities) in red, but do not observe many clear distinctions between core-collapsed and non-core collapsed clusters. The trend in gamma-rays is however slightly stronger when looking only at the core-collapsed clusters, which we note (in advance of the discussion below) are likely to have had much more frequent stellar interactions than their non-core-collapsed counterparts. However, the statistics are still too low to draw any firm conclusion.

The effect of the incomplete sample is quite clear in Fig.~\ref{fig:Luminosities_VX}, particularly for the X-ray case, where we see a large deficit of clusters with small inclinations and lower luminosities.
So to partially mitigate against this bias towards more massive clusters, we now divide out the total dynamical mass, $M$, and look at the \textit{emissivity}, i.e., the $L/M$ ratio---this is shown in the right-hand panels of Fig.~\ref{fig:Luminosities_Gamma} and~\ref{fig:Luminosities_VX}. The emissivity gives us a measure of the cluster's composition rather than its size.

We notice now that the weak correlations in the visible and X-ray luminosities disappear entirely when we consider the emissivity, as would be expected from removing a bias that favours heavier clusters. The act of  However, the correlation between the spin axis inclination and the gamma-ray luminosity persists somewhat. We obtain a 92\% significant exclusion of no correlation that can still be claimed for the gamma-rays, which varies between 85--93\% when one to three clusters are removed from our sample at random. We also note that the cluster that bucked the trend the most when we were only fitting the luminosity, NGC2808, actually matches well the trend fitted by the rest of the clusters when we consider the emissivity. Although the trend is not overwhelmingly significant, it is notable that we should expect those two quantities to be totally random, even if the sample was observationally biased. The fact that there is a discernable hint of a trend in those relations is intriguing.

\section{Further clusters}\label{sec:further}
\begin{figure}
\begin{center}
\includegraphics[trim = 0mm 0mm 0mm 0mm, clip, width=0.49\textwidth]{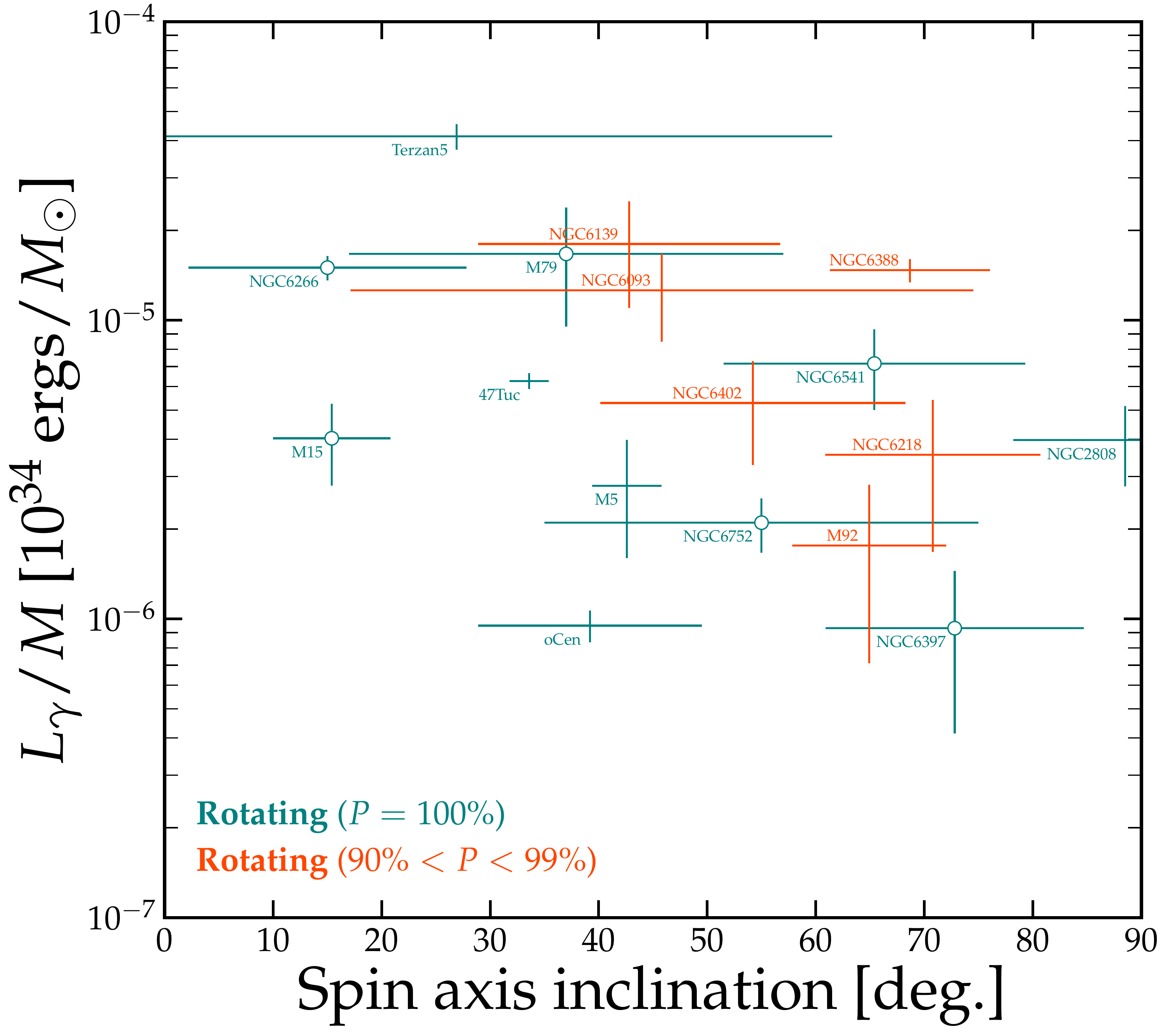}
\caption{Gamma-ray emissivity versus inclination (as in the right-hand panel of Fig.~\ref{fig:Luminosities_Gamma}) but including an additional sample of globular clusters with strong, but not robustly confirmed rotation signals. We estimate the inclination angles using Gaia DR3 proper motions and radial velocities using the method described in Sec.~\ref{sec:further}. These should be treated only as approximate values since the sample of radial velocities is small, and the rotation signal not robustly confirmed, however we do not find any clusters that lie substantially beyond the trend identified previously.} 
\label{fig:extra}
\end{center}
\end{figure} 

The trend between the gamma-ray emission in rotating GCs and the inclination of their spin axes is puzzling and might point towards an interesting physical property. The fact that we do not observe any such trend in the optical or X-ray bands may indicate that it is not solely due to a selection bias effect. To further test this, however, we now consider the objects that may be missing from the strictly selected set of rotating GCs that we have focused on so far.

The recent \textit{Gaia} DR3 included a sample of over 33.8 million line-of-sight velocities obtained spectroscopically~\cite{GaiaDR3_RVs}. These radial velocities, combined with the already large number of stars with proper motions in GCs can be used to estimate their inclination angles arising from a 3-dimensional fit. Assuming a simple model for a solid-body rotation, the stellar velocities in the three dimensions of the cluster relative to the spin axis can be written as:
\begin{equation}
\begin{aligned}
v_Z &=\omega R \sin \left(\theta-\theta_0\right) \sin i \\
v_{\|} &=\omega R \sin \left(\theta-\theta_0\right) \cos i \\
v_{\perp} &=\omega\left[R \cos \left(\theta-\theta_0\right) \cos i+Z \sin i\right]
\end{aligned}
\end{equation}
where the first velocity is the direction pointed along the line of sight, the second velocity is the component in the plane of the sky parallel to the spin axis and the third velocity is the component perpendicular to the spin axis. The position angle of the spin axis in the plane of the sky is $\theta_0$ and the inclination is $i$. The angular velocity $\omega R$ depends on the radius $R$ from the centre,  where $Z$ is the depth of the star inside the cluster along the line-of-sight.

We can obtain an estimate of the rotation axis by combining the amplitude of the rotation signal obtained along the line-of-sight with the amplitude in the plane of the sky. The latter, when projected along the rotation axis can be used to cancel the dependence on the position angle of the rotation axis, to obtain the formula,
\begin{equation}
    i = \arctan{v_Z/v_{\|}} .
\end{equation}
Our estimates using DR3 radial velocities for the clusters in our main sample are all in good agreement with the estimates of~\cite{Sollima_Rotation} within $10^\circ$, with the exception of Terzan 5 for which we have too few stars with line-of-sight velocities.

We include the rest of the clusters with strong rotation signals but only marginally failing the robustness checks in~\cite{Sollima_Rotation}, which we select in this case as those with a probability of rotation between 90 and 99\% (see Table~). We also include clusters identified as rotating in the more recent Gaia EDR3 study of~\cite{Vasiliev_Baumgardt_21_EDR3}. We show these extra clusters in Fig.~\ref{fig:extra}, and list them in Table~\ref{tab:GCs2}. The inclinations are estimated using the above technique, based solely on data from \textit{Gaia} DR3. The error in each case accounts for the uncertainties in each velocity component, the distance, and the systemic motions. Since the number of stars with line-of-sight velocities in DR3 is typically rather small, our corresponding estimate of the position angle of the rotation axis is very poor. This is why we could not include these clusters in the other sections of the paper, we discuss them here solely to determine if they could help explain the apparent correlations as being due to selection bias. We now see that these missing clusters make little difference to our analysis. 

There is still a substantial scatter in Fig.~\ref{fig:extra}, made worse by the relatively poorly measured inclination angles. However, the trend found in the previous section is still visible by eye. So, at the very least, we can claim these clusters missing from the sample are not the cause of bias. 

\begin{figure}
\begin{center}
\includegraphics[trim = 0mm 0mm 0mm 0mm, clip, width=0.49\textwidth]{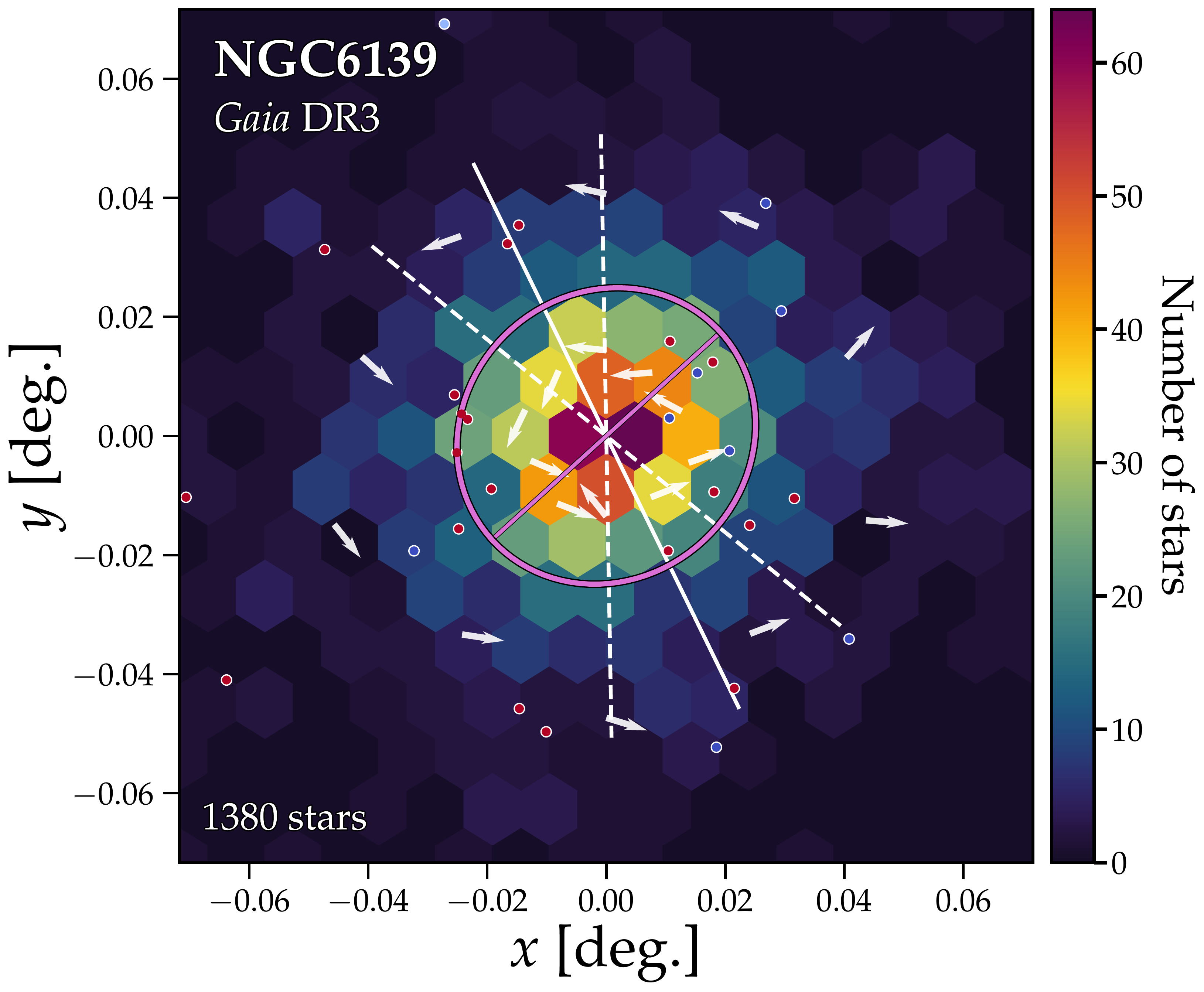}
\caption{Rotation signal for NGC 6139, the only GC with a previously unidentified rotation prior to \textit{Gaia} EDR3~\cite{Vasiliev_Baumgardt_21_EDR3}. We show members with measured line-of-sight velocities as blue/red points, corresponding to negative/positive values relative to the cluster's systemic line-of-sight velocity. The position angle of the cluster measured via star counts is shown by the gold ellipsoid, whereas the position angle measured via the line-of-sight velocities is shown by a white line with $1\sigma$ interval shown as two dashed lines.} 
\label{fig:NGC6139}
\end{center}
\end{figure} 

\begin{table}[t]\centering
\ra{1.3}
\begin{tabularx}{0.49\textwidth}{XX|YY}
\hline \hline
\multicolumn{2}{c}{{\bf Name}} & $N_{\rm stars}$ & $i$ [deg.] \\
\hline
NGC1904 & M79 & 69 & 35.0$^{+18.1}_{-27.6}$ \\
NGC3201 & - & 205 & 72.4$^{+16.8}_{-9.1}$ \\
NGC4372 & - & 503 & 74.0$^{+9.2}_{-8.9}$ \\
NGC5024 & M53 & 40 & 37.9$^{+20.2}_{-30.6}$ \\
NGC5272 & M3 & 138 & 34.6$^{+15.6}_{-18.8}$ \\
NGC5824 & - & 4 & 69.6$^{+23.4}_{-14.8}$ \\
NGC5986 & - & 43 & 59.4$^{+19.6}_{-18.3}$ \\
NGC6093 & M80 & 48 & 45.8$^{+25.8}_{-28.7}$ \\
NGC6139 & - & 34 & 42.8$^{+14.9}_{-13.9}$ \\
NGC6218 & M12 & 128 & 70.8$^{+14.5}_{-9.9}$ \\
NGC6341 & M92 & 103 & 64.9$^{+7.3}_{-7.1}$ \\
NGC6388 & - & 76 & 68.7$^{+7.3}_{-7.4}$ \\
NGC6402 & M14 & 88 & 54.2$^{+14.5}_{-14.1}$ \\
NGC6496 & - & 29 & 74.2$^{+22.0}_{-11.3}$ \\
NGC6539 & - & 31 & 60.2$^{+24.1}_{-19.5}$ \\
NGC6752 & - & 754 & 43.6$^{+19.6}_{-15.1}$ \\
NGC6809 & M55 & 166 & 73.4$^{+10.9}_{-7.0}$ \\
NGC7099 & M30 & 44 & 77.9$^{+15.0}_{-8.5}$ \\
\hline \hline
\end{tabularx}
\caption{Estimated inclination angles for a further sample of clusters, and the number of members in Gaia DR3. We do not include clusters with fewer than 30 measured radial velocities in Fig.~\ref{fig:extra}. \label{tab:GCs2}}
\end{table}

One interesting case is the cluster NGC 6139 which had no recorded measurements of rotation prior to the release of Gaia DR3~\cite{Vasiliev_Baumgardt_21_EDR3} but has rotation in the plane of the sky that is apparent even by eye, as can be seen in Fig.~\ref{fig:NGC6139}. The amplitude of the oscillation in the line-of-sight, $v_{\rm los}(\theta)$, and proper motion projected along the rotation axis, $v_{\\}(\theta)$, are roughly the same, hinting that the cluster is likely oriented in such a way that it generates a similar rotation signal both in the plane of the sky and the line-of-sight, i.e. $i \sim 45^\circ$. Using the method above, we obtain $i = 43^\circ$, which to our knowledge, is the first reported inclination angle for this cluster. To lend further confidence to this result, we can use an alternative technique described in~\cite{Bianchini_RotatingDR2} wherein one takes the ratio of the proper motions along the major and minor axes, i.e.
 \begin{equation}
     i \approx \cos^{-1}\bigg( \frac{v_t(\phi)}{v_t(\phi+90^\circ)} \bigg) \, ,
 \end{equation}
 where $v_t(\theta)$ is the tangential proper motion as a function of angle around the cluster, and $\phi$ is the position angle of the cluster's rotation axis obtained separately. Comparing the tangential motions in two orthogonal angular bins of size 5--15$^\circ$ for NGC6139, we obtain $i \approx 40^\circ$ in good agreement with our other estimate. Since the gamma-ray luminosity for this cluster is also reasonably high, $L_\gamma = 5.28 \times 10^{34}$ erg s$^{-1}$, this cluster therefore warrants further analysis to pin down its kinematics more precisely.

Further clusters that have been measured with strong gamma-ray emissivities above $10^{-5}$ ($\times 10^{34}$ erg s$^{-1} \ M_\odot^{-1}$) are: Terzan 1, Terzan 2, NGC6440, NGC6441, NGC6717, NGC6528, and NGC6652. Of these NGC 6640 and NGC 6441 have tentative but very low significance rotation signals, and only NGC 6441 has a decent sample in DR3 ($\sim$220). The emissivity of the latter cluster is $1.3\times10^{-5}$ and we estimate an inclination angle of $\sim50^\circ$. 

\begin{figure}
\begin{center}
\includegraphics[trim = 0mm 0mm 0mm 0mm, clip, width=0.49\textwidth]{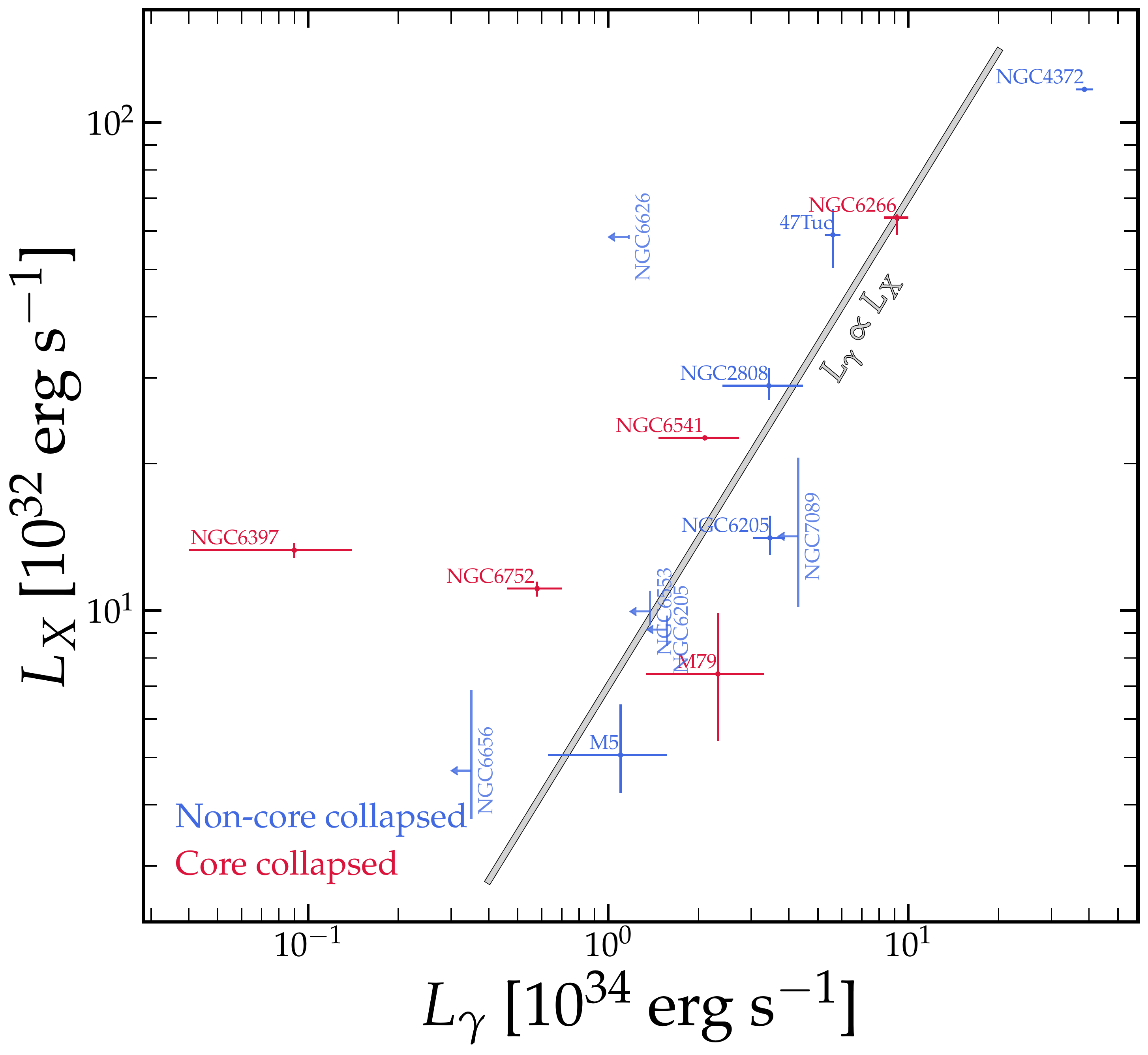}
\caption{Gamma-ray luminosity versus X-ray luminosity. The straight line indicates the proportionality $L_\gamma \propto L_X$. GCs with only upper limits on their gamma-ray luminosity are shown with arrows. We make no statement about the origin or robustness of this trend, only that we wish to identify plausible values for the gamma-ray luminosity in those rotating clusters that only have upper limits.} 
\label{fig:Gamma_vs_Xray}
\end{center}
\end{figure} 

As a final test for the persistence of this correlation in the gamma-ray emissivity and inclination, we can discuss the clusters in our rotating sample that only have 95\% upper limits on $L_\gamma$. These are shown by the transparent lines with arrows in Fig.~\ref{fig:Luminosities_Gamma} (lower right panel). If we take the most conservative approach and assume the true values are at the edge of their upper limits, our exclusion of a gradient $>0$ increases to 95\%. If, instead, we adopt a more optimistic approach of taking the value of $L_\gamma$ that correlates with the $L_X \propto L_\gamma$ trend shown in Fig.~\ref{fig:Gamma_vs_Xray}, then the exclusion increases to 99\% CL. This suggests that a clearer picture of this correlation could be obtained with further analysis of these six clusters.

\begin{figure}
\begin{center}
\includegraphics[trim = 0mm 0mm 0mm 0mm, clip, width=0.49\textwidth]{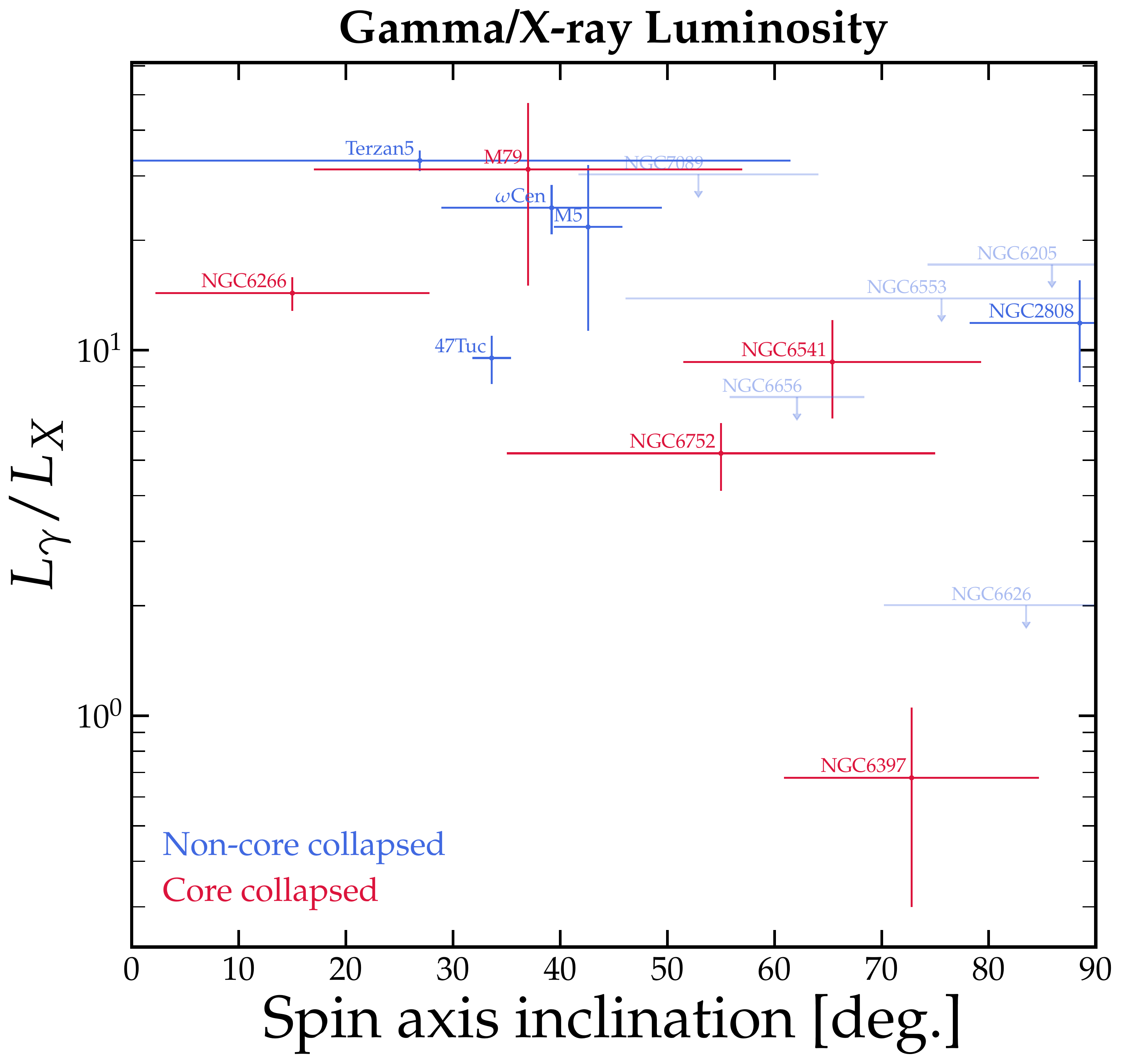}
\caption{Ratio between the gamma-ray and X-ray luminosities versus the inclination angle. Again we see here evidence of a negative correlation, despite the fact that a purely selection-bias driven effect would be expected to be removed when taking this ratio. It appears there are certain clusters like NGC6397 that have a gamma-ray luminosity that is anomalously low compared to its X-ray luminosity---an observation that could be partially explained due to anisotropic emission effect in the gamma-rays which can be accounted for by considering the cluster's orientation.} 
\label{fig:Gamma_ovr_Xray}
\end{center}
\end{figure}

Lingering for a moment on the X-ray versus gamma-ray phenomenology of GCs as illustrated by \autoref{fig:Gamma_vs_Xray}, we note that clusters like NGC 6397 that fall well to the left of the fit line tend to be highly inclined. This can be shown in a plot of the ratio of a GC's apparent gamma-ray luminosity to its X-ray luminosity versus its spin axis inclination which we show in \autoref{fig:Gamma_ovr_Xray}. As expected, this ratio is anti-correlated with $i$. This is not new information providing independent evidence for a physical effect underlying the anti-correlation of the apparent gamma-ray luminosity with $i$, but it does suggest that such an interpretation may be consistent, as we shall now explore.

\section{A physical explanation for anisotropic gamma-ray emission in globular clusters}\label{sec:physicalexplanation}

We have identified an intriguing hint from a sub-sample
of Milky Way globular clusters that their apparent gamma-ray luminosities have an anti-correlation with the inclination of their spin axes with respect to our line of sight, $i$. To our knowledge, this anti-correlation was not theoretically predicted; it was certainly a surprise to the authors. It is possible that the apparent anti-correlation is due to some selection bias. We have, however, attempted to test this by searching for correlations between the emissivity (or mass-to-light-ratio) measured in various wavebands (visible, X-ray and gamma-rays) and the inclination angle, and by seeking out clusters that the tight selection efficiency on rotating clusters may have been biased against. While the correlation with the gamma-rays survives these tests, apparent correlations with the X-rays and visible light, already very weak or non-existent, do not. Additionally, it seems that there are no obvious samples of rotating clusters that could be missing from the sample, and thus induce a correlation between otherwise uncorrelated variables by their absence.

Given the statistical evidence is not yet definitive, it is also conceivable that the apparent anti-correlation is a statistical fluctuation. As for selection bias, however, we tried to press on this above by increasing the sample size by including GCs with relatively more poorly constrained kinematics and/or those GC without a firm gamma-ray detection in~\cite{Song_Fermi_GCs}, but rather only an upper limit. As we have already seen, by filling out the sample size like this, the statistical strength of the anti-correlation is not reduced.

We are, therefore, motivated to consider---in a somewhat speculative mode---the idea that the explanation for the anti-correlation is that globular clusters emit gamma-rays anisotropically, preferentially towards the poles of their rotation axes.

A physical explanation for this observation begs for an anisotropic source of gamma-rays inside globular clusters. In particular, it demands the existence of sources that themselves emit anisotropically and which also tend to be aligned with the overall rotation axis of the GCs.

As presaged above, there is very good evidence millisecond pulsars are the overwhelming source of gamma-ray emission from GCs \citep{Chen91, Abdo09_47Tuc_Gammas,Abdo10} with such evidence now including the detection of phase-resolved, pulsed gamma-ray emission from a limited number of them~\citep{Freire2011, Wu2013}. Given this, we are driven, almost inevitably, to the conclusion that any anisotropic emission from GCs should somehow involve MSPs.

Two radiation mechanisms---curvature radiation from individual pulsar magnetospheres and inverse Compton (IC) radiation from (MSP-escaped) $e^\pm$ off ISM lightfields---likely contribute to the overall gamma-ray signal from GCs \citep{Bednarek2007, Venter2009, Cheng2010, Song_Fermi_GCs}. Thus, in principle, only one or the other might be responsible for the angular dependence of radiant intensity. 

However, we have not been able to arrive at a satisfactory physical explanation invoking IC that is a match to the overall phenomenology that we explore above\footnote{An IC explanation would seem to require a large-scale, coherent magnetic field structure with field lines tending to extend along the rotation axes of the cluster and, moreover, a population of IC-emitting $e^\pm$ pairs that are not pitch angle isotropised; we cannot arrive at a scenario that naturally satisfies these requirements, but certainly do not discourage others from exploring a mechanism along these lines.}. In any case, previous work has found that IC from GCs is sub-dominant with respect to curvature radiation in the Fermi band~\cite[e.g.,][]{Song_Fermi_GCs} and it seems a stretch that, given this, even a 100\% variation in the IC radiant intensity over $i: 0 \to 90^\circ$ could account for the amplitude of the angular dependence. Thus, a physical explanation for the anti-correlation seems necessarily to be related to the curvature radiation, as we will explore below.

We note in passing that, if, as expected, IC emission from GCs is isotropic, whereas curvature radiation tends to be emitted along the rotation axis, then we expect a systematic change in the detected spectrum of the \textit{Fermi}-band gamma-ray emission from GCs as a function of $i$. In particular, we expect that, for clusters with small values of $i$, the `GeV bump' in the SED due to curvature radiation should be more apparent, while for $i \to 90^\circ$, the spectrum should become more power-law-like due to a relatively larger contribution of IC. In Appendix~\ref{sec:gamma-ray-spectrum} we explore the possibility of just such a correlation between the inclination angle and the relative contribution to the gamma-ray luminosity from each mechanism as determined by~\cite{Song_Fermi_GCs}; unfortunately, at present the results of this exercise are inconclusive.

To summarise the developments above, a physical explanation for the anti-correlation must somehow involve the summed, magnetospheric emission from the MSPs in a GC. To self-consistently reproduce the anisotropy, it is then required that i) the magnetospheric gamma-ray emission from individual MSPs is anisotropic; ii) the direction of brightest emission from individual MSPs tends to be aligned at a population level; and iii) the overall direction along which the brightest emission from individual MSPs tends to align, tends itself to be close to the spin axis of the host GC.

Of these conditions, the first seems to be directly and naturally satisfied, at least at the qualitative level: magnetospheric gamma-ray emission from MSPs is certainly anisotropic \citep[e.g.,][]{Abdo13}.
Indeed, the preferential emission of non-thermal radiation 
in beam patterns shaped by the magnetosphere---where such beams are, in general, misaligned with the spin axes---explains the defining periodicity of pulsar and MSP lightcurves, i.e., the eponymous pulses, at radio and/or gamma-ray wavebands.


The next requirement for a physical origin for the anti-correlation with $i$--- 
that the directions of brightest gamma-ray emission from individual MSPs tend to be aligned at a population level---seems to imply 
one of two situations:  
\begin{enumerate}
\item A ``perpendicular-perpendicular" geometry wherein individual GC MSPs' spin axes tend to lie in the plane perpendicular to the host GC's rotation axis while the gamma-ray beams of these MSPS tend, in turn, to be perpendicular to their spin axes. In other words, in this scenario, an MSP's magnetic inclination angle $\alpha $---the angle between its spin and dipole magnetic field axes---should be approximately $\pi/2$. To see how this scenario works at the qualitative level to achieve the $i$-dependence of the apparent gamma-ray luminosity, consider the limit of a large MSP population with spin axes filling the azimuthal angle around the equatorial plane of a GC's rotation axis, small beam opening angles, and perepndicular magnetic inclination angles. In this configuration, the gamma-ray beam of every MSP sweeps through the poles of the GC's rotation axis, but only a small fraction of all MSP beams will sweep past an observer (located at any azimuthal angle) in the equatorial plane of the GC's rotation axis.

\item A ``parallel-parallel'' geometry wherein individual GC MSPs' spin axes tend to be parallel to the host GC's rotation axis while their gamma-ray beams tend to also be close to parallel to their spin, i.e.~the MSPs' magnetic inclination angles $\alpha$ should also be small.
\end{enumerate}

As we now explain, the radio phenomenology of MSP detections in GCs seems to indicate that the parallel-parallel scenario is preferred over the perpendicular-perpendicular. This follows from the fact that the scenarios have different predictions for the $i$-dependence of the ease of detecting individual MSPs within GCs, something which is usually accomplished via radio observations.

More precisely, they predict a different $i$-scaling of the ratio of the number of detected MSPs, $N_{\rm MSP}$, and the apparent gamma-ray luminosity. Consider the following: for a pulsar to be positively identified as such in the radio band (or, indeed, in the gamma-ray band), pulsed emission must be identified in its light curve. In the perpendicular-perpendicular geometry, any MSP that is visible to an observer will (in principle) pulse. As already mentioned, in this geometry the beam of every MSP in the GC sweeps over an observer looking down the host's rotation axis. Observers with this position should detect both the strongest (time-integrated) apparent gamma-ray luminosity and the largest number of pulsing MSPs (in principle, they should detect every MSP in the GC with an individual power above some detection threshold).

On the other hand, in the parallel-parallel geometry, the beams of every MSPs always point at an observer who happens to be located down the GC's rotation poles; so, again, the GC has the greatest apparent gamma-ray luminosity from this privileged position but, on the other hand, in the limit of small magnetic inclination angles, there are no pulsations because each MSP's beam is always pointing at that observer. In other words, from the position of this observer, individual MSPs cannot be identified (at least in light curve data).

Relaxing the various limits appealed to above, we have a qualitative expectation that, for the perpendicular-perpendicular case, the ratio $N_{\rm MSP}/L_\gamma$ is a declining function of $i$, while for 
parallel-parallel it is not.

%


%

To investigate this, we source numbers of identified, individual pulsars in GCs from a public database\footnote{\url{https://www.naic.edu/~pfreire/GCpsr.html}} and in Fig.~\ref{fig:NumberOfPulsars_inclination} we show the ratio of the number of pulsars to the apparent gamma-ray luminosity as a function of the inclination angle. The result is a very clear positive trend, consistent with expectation for the parallel-parallel scenario (where, if anything, we expect pulsations to become easier to identify at high inclination angles). 

A word of caution with respect to \autoref{fig:NumberOfPulsars_inclination} is that there are three clusters in this sample with measured $L_\gamma$ and inclination, but have zero detected pulsars (M79, NGC6541, NGC2808). These are all some of the most distant clusters in the sample, however, this is a reminder that the quantity $N_{\rm MSP}$ will come with its own set of selection biases, which are harder to untangle than the other quantities we have discussed.

We also note that the number of detected pulsars is not correlated at all with the inclination angle, but \textit{is} correlated relatively strongly (faster than linear) with the gamma-ray luminosity. It is possible then that this correlation we observe here is due to the strong dependence of $N_{\rm MSP}$ on $L_\gamma$, which leads to a net positive correlation when the ratio of the two is plotted against $i$. Nevertheless, given that the reverse trend is expected in the perpendicular-perpendicular scenario, this is enough to cause us to disfavour it over the alternative\footnote{We thank Simon Johnston for pointing out a third possible geometry that is consistent with the apparent GC inclination angle dependence of the apparent gamma-ray luminosity and with \autoref{fig:NumberOfPulsars_inclination}: if we do not require that MSPs' gamma-ray and radio beams overlap but, rather, allow for a configuration where radio emission is beamed close to the spin axes of individual MSPs, while they also tend to emit gamma-rays in the equatorial plane defined by their spin axes, then, in the case that the spin axes of MSPs in a GC tend to be perpendicular to the GC's rotation axis, we seem to qualitatively account for the overall 
phenomenology. We do not further consider this case because, 
by construction, it is in tension with the population statistics of radio vs.~gamma-ray discovery of individual MSPs 
which tend to suggest that their 
"gamma-ray beams cover a comparably-sized, and nearly-coincident, fraction of the sky"
\citep[ref.][section 10.1]{Abdo13}. 
}.

\begin{figure}
\begin{center}
\includegraphics[trim = 0mm 0mm 0mm 0mm, clip, width=0.49\textwidth]{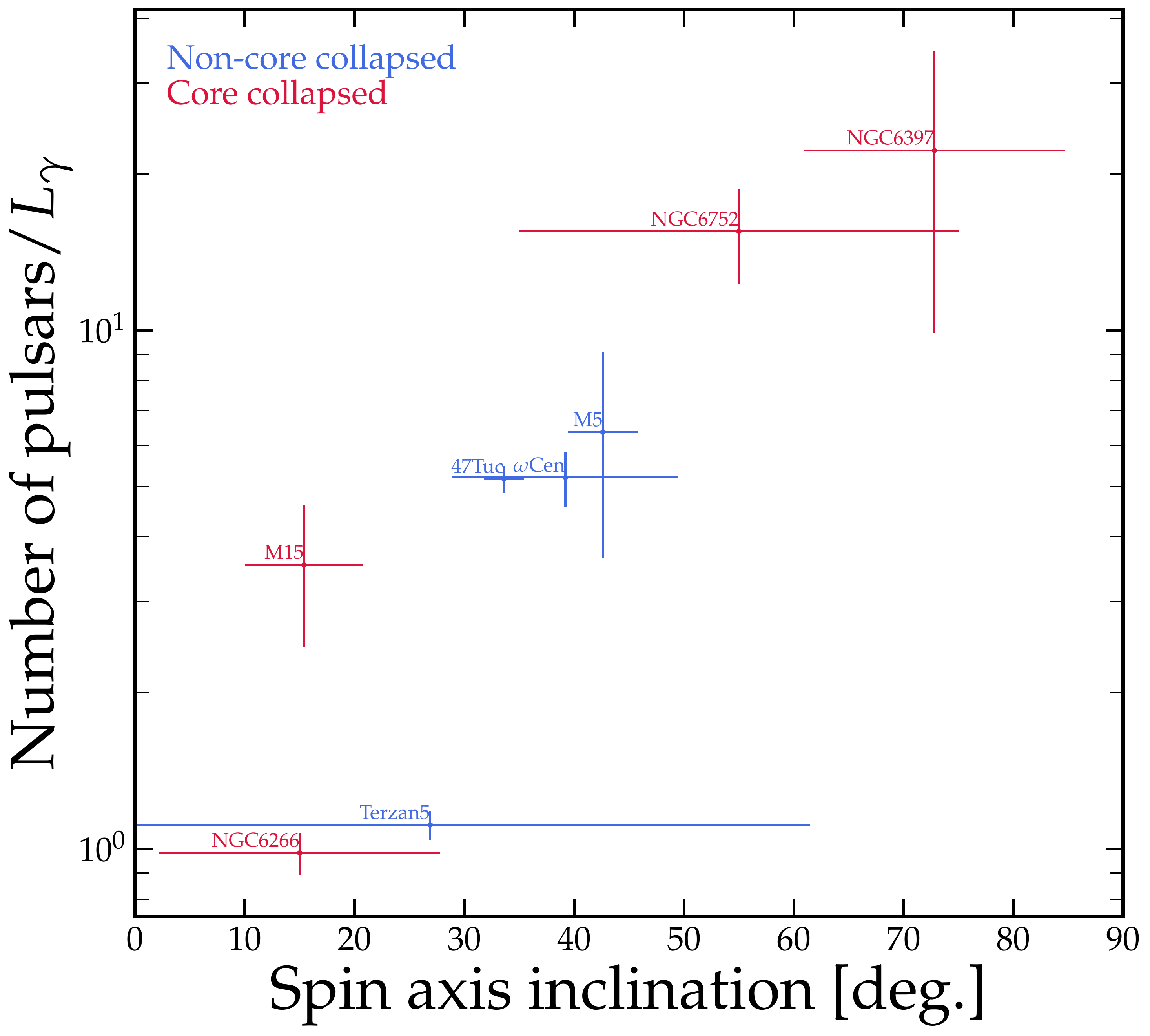}
\caption{Number of identified pulsars divided by the gamma-ray luminosity, as a function of the spin-axis inclination.} 
\label{fig:NumberOfPulsars_inclination}
\end{center}
\end{figure}

In light of these considerations, we are left with the following broad physical scenario underlying the anti-correlation with $i$: we require

\begin{enumerate}
\item that the
spin axes of individual GC MSPs tend to be aligned, and aligned, moreover, with the rotation axis of the host GC;
\item that the mean MSP magnetic inclination angle $\alpha$ is not very large;
\item that the \textit{dispersion} in MSP magnetic inclination angle is not very large; and
\item that the typical MSP gamma-ray beam is not very wide or, in the likely~\cite[e.g.,][]{Abdo13} case that it is wide, there is still a sufficiently strong angular variation of the radiant intensity within the beam solid angle that the anticorrelation can be accommodated.
\end{enumerate}

It seems non-trivial to match all these requirements with GC MSP populations.
Furthermore, some of these requirements actually translate into quantitative (but degenerate) restrictions on, e.g.~the gamma-ray beam geometry, which we have not investigated here. On the other hand, we do note that at least
some of those points seem quite plausible.

In particular, given that the spin-down  power emitted by an inclined magnetic dipole scales like $\sin^2\hspace{-0.05cm}\alpha$, for a sufficiently old MSP population (as may, indeed, be characteristic of a GC), one expects that the fastest remnant rotators are the ones with the smallest $\alpha$ values (either because this subset were created with small $\alpha$ or because their $\alpha$ values evolved to be small). This same $\alpha$-dependence to the magnetic braking ensures that there is a small dispersion in the $\alpha$ after sufficient time has passed. Furthermore, assuming, for the moment, that there does exist a clear direction along which the spin angular momenta of a GC's MSPs tends to align, it seems reasonable that this would be the GC's own rotation axis (in the absence of any other compelling candidate for a special direction).

That said, it is not obvious that information about the cluster-scale angular momentum direction could be reliably communicated down to the scale of individual MSPs. A particular consideration here is that GCs exhibit substantially higher gamma-ray efficiency (per unit stellar mass) than field star populations~\citep[e.g.,][]{deMenezes:2018ilq,Fermi_GCs,Crocker2022}. This has been interpreted as indicating a higher efficiency for forming MSPs (per unit mass of all stars formed), presumably as a result of dynamical processes that increase the overall prevalence of close binaries
or tend to harden some of them. 

It is perhaps conceivable that, in a rotating system, stellar encounters forming or hardening binary systems might leave them with orbital planes aligning with the overall plane of rotation (or, if dynamical encounters within GCs tend to wash out the required alignments, perhaps secular processes can restore it). In any case, we must recall that MSPs are formed via accretion (or merger) through the orbital plane in close binary systems\footnote{MSPs are products of close and interacting binaries wherein either an existing neutron star, born of a previous core-collapse supernova event, is spun up via accretion from a binary companion \citep{Backus1982,Radhakrishnan1982} in LMXB-like episodes or a massive O-Ne white dwarf, again accreting from a close companion and approaching the Chandrasekhar limit, experiences accretion induced collapse (AIC) down to a rapidly-rotating neutron star \citep{Miyaji1980,Ruiter2019,Gautam2022} or, finally, they may be produced by Merger Induced Collapse (MIC) following the coalescence of a O-Ne white dwarf with a white dwarf companion \citep[e.g.,][]{Ruiter2019}; note that MIC will also produce a final neutron star whose spin is in the plane of the natal coalescing binary.} where the final spin angular momentum of the `spun-up' neutron star should, then, be roughly parallel to the overall orbital angular momentum of the binary. Therefore, information about the overall cluster angular momentum needs, in principle, only to be communicated down to the size scale of binary systems, rather than individual stellar objects. 



While we admittedly have no detailed account of how the required ``spin-orbit coupling" happens, there are some observational hints for the existence of this kind of effect in stellar clusters, in particular, for open clusters~\cite{Corsaro17_SpinAlignmentOpenClusters,Kovacs2018_praesepe}. Most notably Corsaro et al.~\cite{Corsaro17_SpinAlignmentOpenClusters}, used asteroseismology to measure the spins of stars in two old open clusters, NGC 6791 and NGC 6819, finding evidence of strong stellar spin alignment.

Furthermore, the study of binaries in~\cite{Klagyivik2013_binaryalignment}, suggested that they were aligned in the inner dense region of a young cluster.  However, it should also be highlighted that a more recent study of open clusters was consistent with an isotropic alignment~\cite{Healy_openclusterspins}. In the case of globular clusters, on the other hand, the effect of the spin transfer to the stars has been investigated ~\cite[e.g.][]{Kamann_spinorbit}, but only suggestive evidence was found for spin-orbit alignment in NGC6791, and no evidence in NGC6819.

On the modelling side, the simulations of~\cite{ReyRaposo2018_spinalignementsimulation} report that the initial angular momentum of a natal molecular cloud could be inherited and preserved by the stars it forms via their individual spin angular momenta implying spin alignments that survive through to late times.

In summary here, the gamma-ray indication we have found that individual GC MSPs retain, in their spin orientation, a memory of their host's overall rotation axis, is certainly surprising but, to our knowledge, is not trivially ruled out. Indeed, there are (contested) observational indications, as summarised above, for a similar or identical effect operating in open clusters.

\section{Discussion and conclusions}
\label{sec:discussion}
In this study, we have considered a sample of Milky Way Globular clusters (GCs) with robustly measured rotation signals, that allow for 3-dimensional rotation axes to be measured and oriented in the Galaxy. We have investigated the broad question of whether or not these spin axis directions are correlated with other properties of the cluster. While there are the beginnings of some potentially interesting hints to do with the orientation of the spins for clusters brought into the Milky Way via different merger events, so far the size of the sample is not sufficient for a significant claim to be made contrary to the expectation of isotropically distributed spins. 

However, when we consider the relationship between the spin axis orientation compared to the line of sight and the apparent luminosities of the clusters we find an intriguing negative correlation at gamma ray wavelengths. Given that GCs are 
almost perfectly spherically symmetric agglomerations of stars, this observation is extremely surprising. 
In fact, it is highly suspicious given there could be sources of selection bias.
It is easier to measure the 3d spin axes orientations of GCs that are rotating strongly, whilst the extraction of spins requires both plane-of-the-sky and line-of-sight rotation signals which are subject to differing levels of systematic uncertainty.
However, as far as we have been able to tell, this selection bias effect cannot conclusively 
explain the observed signal in the gamma-rays. This is because the induced correlation between the luminosity in the X-ray and optical bands can be removed when dividing out the cluster's dynamical mass, which, as Fig.~\ref{fig:Luminosities_Gamma} shows, does not remove the correlation in the case of the gamma-ray luminosity.

We are left then, to contemplate the possibility that GCs are truly anisotropic emitters, but only in gamma-rays. Given that these gamma rays 
are strongly believed to arise 
from populations of millisecond pulsars (MSPs)---which, individually, are anisotropic gamma-ray emitters---such a hypothesis, while still demanding several factors to work together, is not trivially ruled out, as it would be in other wavelengths.

\subsection*{Next steps}

In Section~\ref{sec:physicalexplanation}, we summarised the various features that the MSPs must have at both the population and individual level for GCs to emit gamma-rays anisotropically. So to round off this discussion we will now recommend several further steps that should be made next to determine whether the correlation we have uncovered is a physical one.

First of all, a larger and more precise sample of line-of-sight spin-axis inclination angles should be obtained using astrometric and spectroscopic data from the upcoming Gaia Data Releases and proposed missions as JASMINE \citep{Gouda:2011}, \textit{Gaia}NIR \citep{2016arXiv160907325H}, \textit{Theia} \citep{2017arXiv170701348T}, in combination with HST, JWST, NGRST and other ground-based observatories. 

Secondly, another obvious place to push is on the number of gamma-ray-detected GCs. Indeed, as already mentioned in \autoref{sec:further}, the positive detection of 6 GCs for which only upper limits are reported in~\cite{Song_Fermi_GCs} (NGC4372, NGC6205, NGC6273, NGC6553, NGC6626, NGC6656, NGC7089) holds out the prospect of securing 
the existence of the gamma-ray anti-correlation with $i$ at the 99 \% CL. In general, improving the size of the sample and its precision should make it clearer if there are sampling biases still at play. 

Thirdly, if the correlation persists in the gamma-ray luminosities, it should also be tested in other wavelengths, in particular, the radio and hard (non-thermal) X-ray to further test for a millisecond pulsar origin. 

%

We can make some other predictions about the phenomenology of MSPs in GCs; no doubt some of these are less observationally tractable than others, but any of them would constitute broadly independent evidence confirmatory of our tentative physical interpretation of the gamma-ray anti-correlation with $i$:
\begin{enumerate}
\item To the extent that the summed magnetospheric emission from all MSPs in a GC dominates its unresolved, non-thermal $\sim$GHz band radio continuum emission, then we would expect this unresolved emission, like the gamma-ray emission, to be anticorrelated with $i$. In particular, the unresolved, total flux density ascribable to GCs (assuming this is measurable) should exhibit an anti-correlation with $i$.
\item The spin axis orientations of MSPs in any one cluster that can be inferred from modelling of individual MSP's radio light curves should tend to be aligned both between themselves and with the Gaia-inferred $i$ of the GC host.
\item The inclination of the  orbital planes of MSP-containing
binaries in GCs should also tend to align both between themselves and with the host GC's inclination angle. This might be investigated with the sort of techniques discussed in refs.~\cite{Guillemot2014, Smedley2014, Smedley2015} with respect to the orbit modelling of MSPs with He white dwarf binary companion.
\item The previous point suggests that, at least in principle, gamma-ray eclipses due to the binary companion in MSP `spider' systems
\cite[cf.,][]{Clark2023} should, in principle, be more frequent in high $i$ clusters than low $i$ ones.
\item The ratio of the apparent gamma-ray luminosity of a GC to the sum of the 
spin-down power delivered by all its MSPs (as inferred by radio measurements of the MSPs' period and period derivatives with the latter corrected for acceleration due to the GC's gravitational potential) should be a declining function of $i$.
\end{enumerate}

As we have already emphasised, confirmation of a physical mechanism
underlying the gamma-ray anticorrelation with $i$ 
would have important implications for the geometries of the beams and magnetic inclination angle distributions of GC MSPs. In particular, it would seem that the phenomenology we have uncovered favours magnetospheric emission scenarios wherein MSPs emit most strongly along a direction close to their spin axes rather than perpendicular to their spin axes. It also seems necessary that either the solid angles of the beams of GC MSPs are smaller than seem to be indicated by other data or that,  even if ``wide", they must still accommodate a fairly large diminution in radiant intensity between the beam centre and edge.

Finally, also as already emphasised, were the gamma-ray anti-correlation with $i$ a real physical effect,  this would constrain the dynamical evolution of dense, gravitationally bound star clusters in interesting ways. Indeed, it seems warranted to suggest deeper investigation of the potential mechanisms behind the sort of ``spin-orbit" coupling requisite to explain the anti-correlation via N-body or other modelling schemes. It is particularly desirable to test for the timescales over which such an anisotropy could arise and be maintained, and determine whether or not it is reasonable for a strong present-day correlation to be observed. We hope to inspire future research along exactly these lines.

\appendix
\section{Cluster rotation continued}
In Fig.~\ref{fig:Streamline_cont}, we show the rest of our sample of rotating clusters as we did in Fig.~\ref{fig:Streamline}. We neglect to show clusters with fewer than 1000 stars where the rotation signal on the plane of the sky is less apparent by eye.
\begin{figure*}
\begin{center}
\includegraphics[trim = 0mm 0mm 0mm 0mm, clip, width=0.33\textwidth]{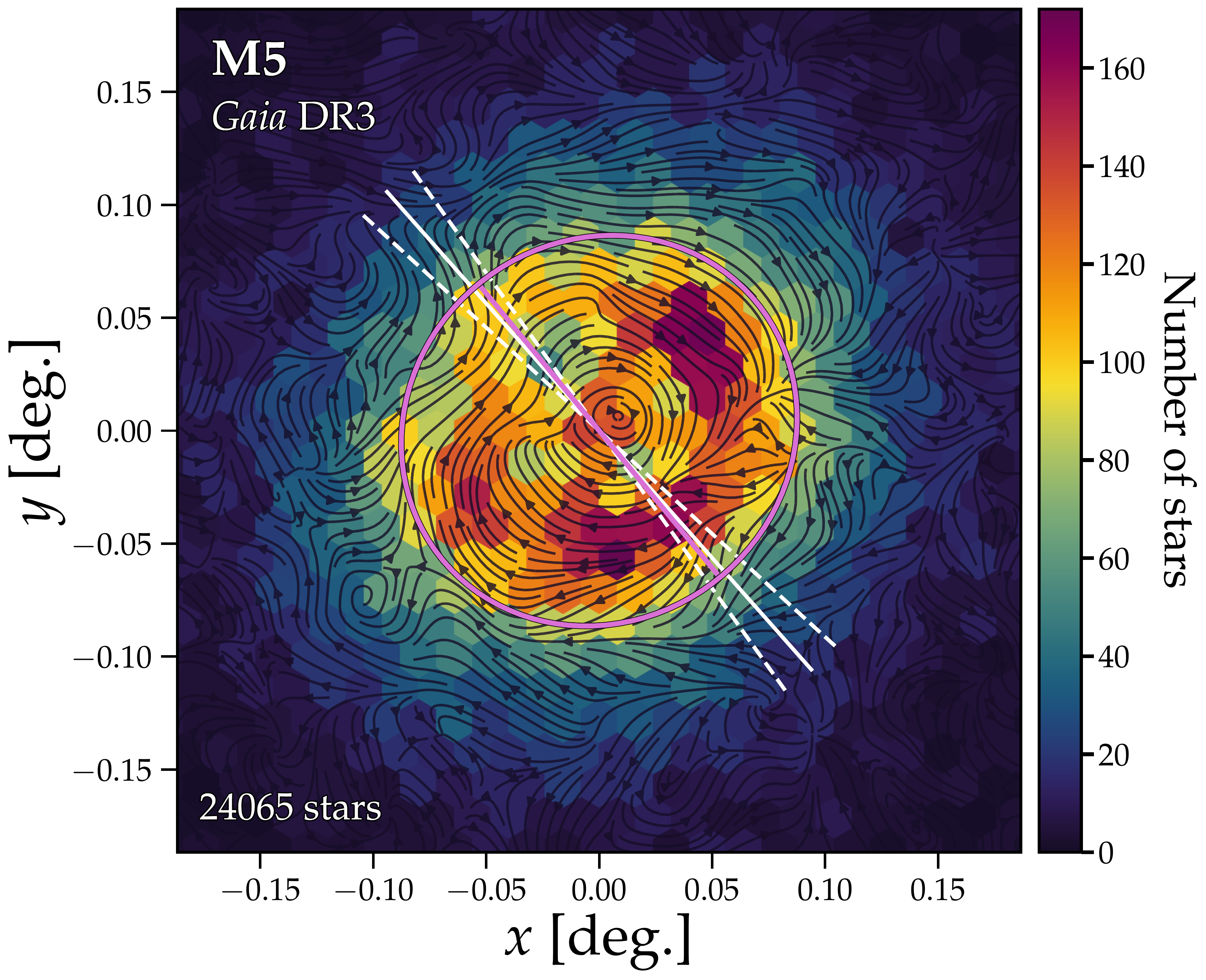}
\includegraphics[trim = 0mm 0mm 0mm 0mm, clip, width=0.33\textwidth]{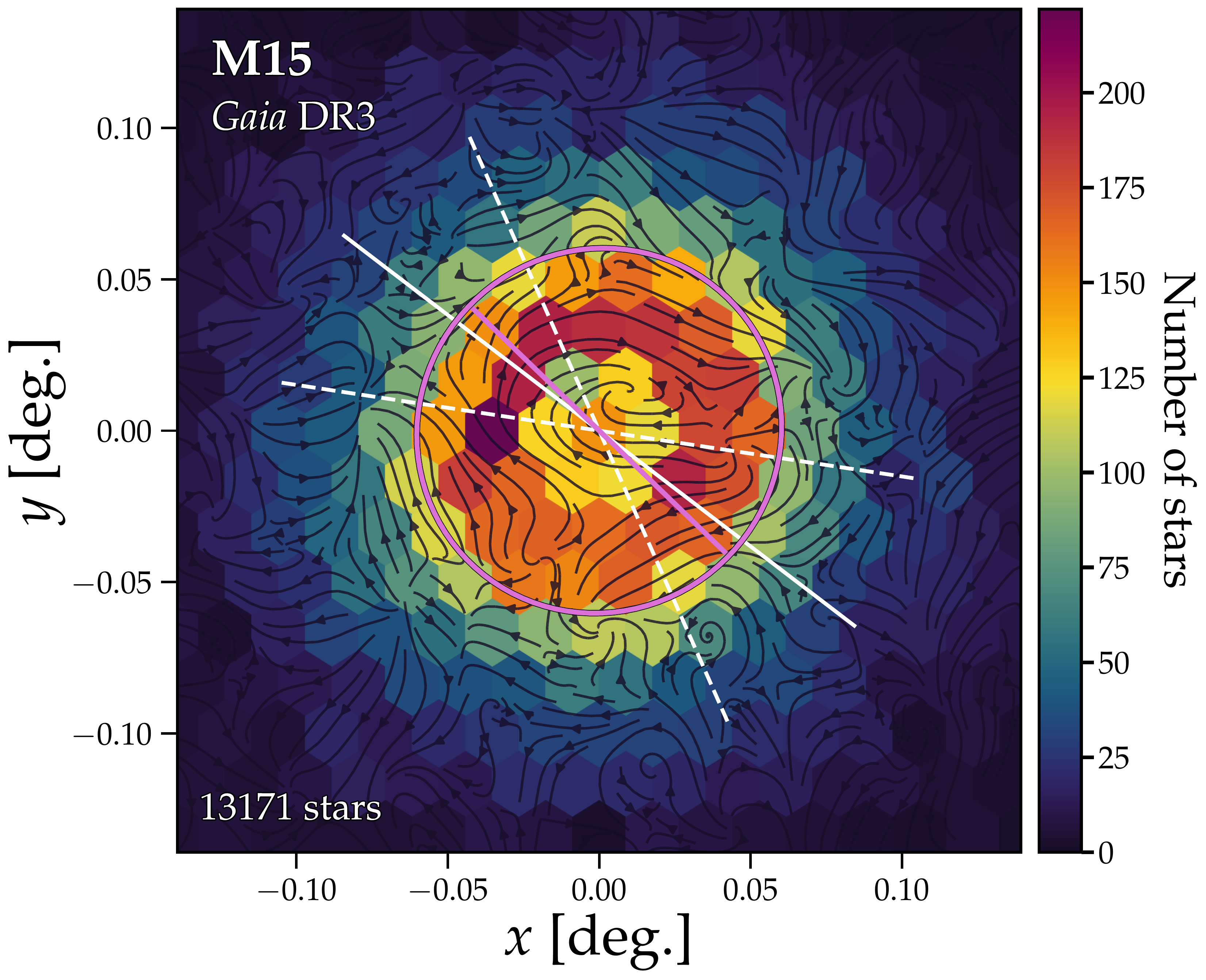}
\includegraphics[trim = 0mm 0mm 0mm 0mm, clip, width=0.33\textwidth]{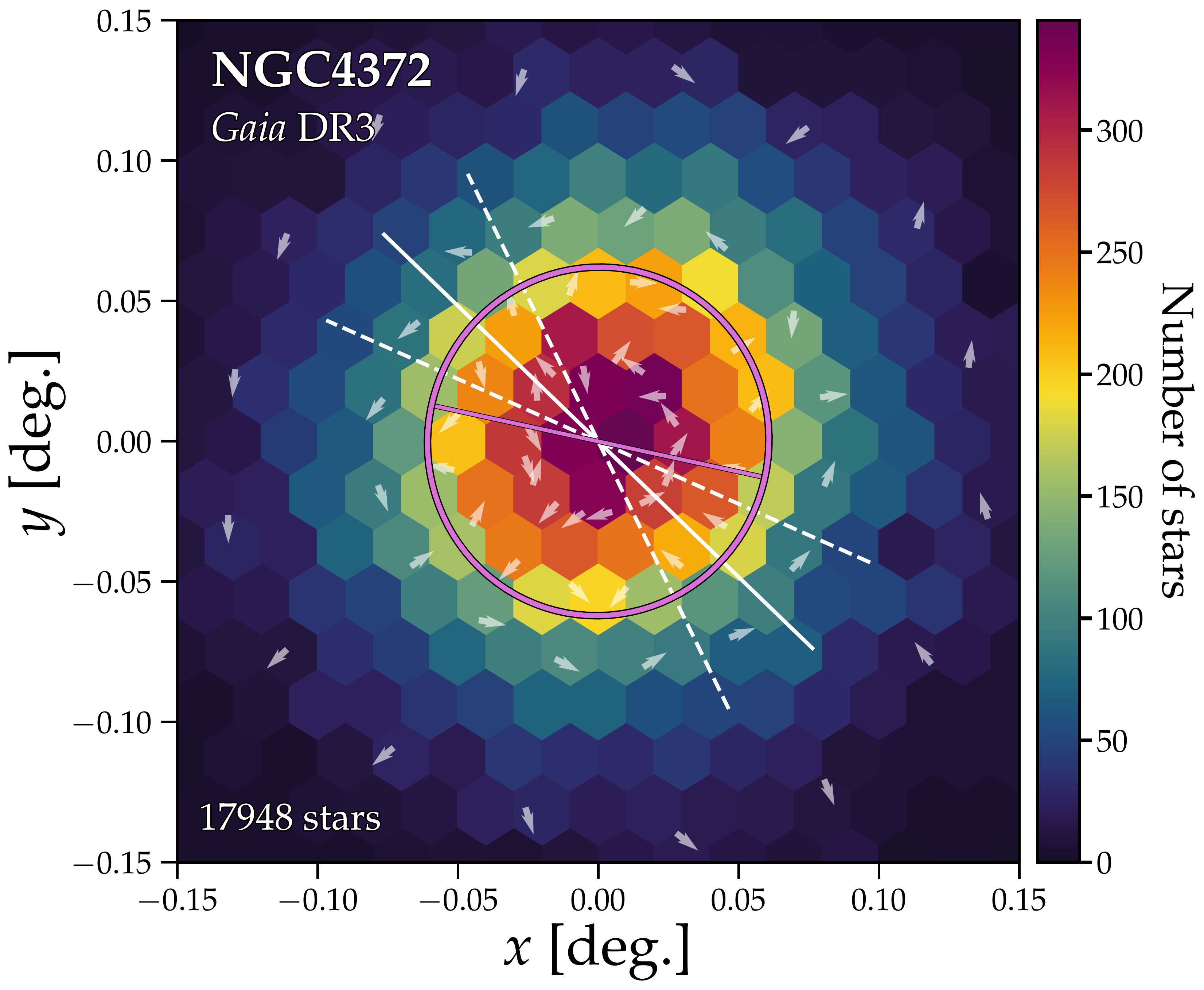}
\includegraphics[trim = 0mm 0mm 0mm 0mm, clip, width=0.33\textwidth]{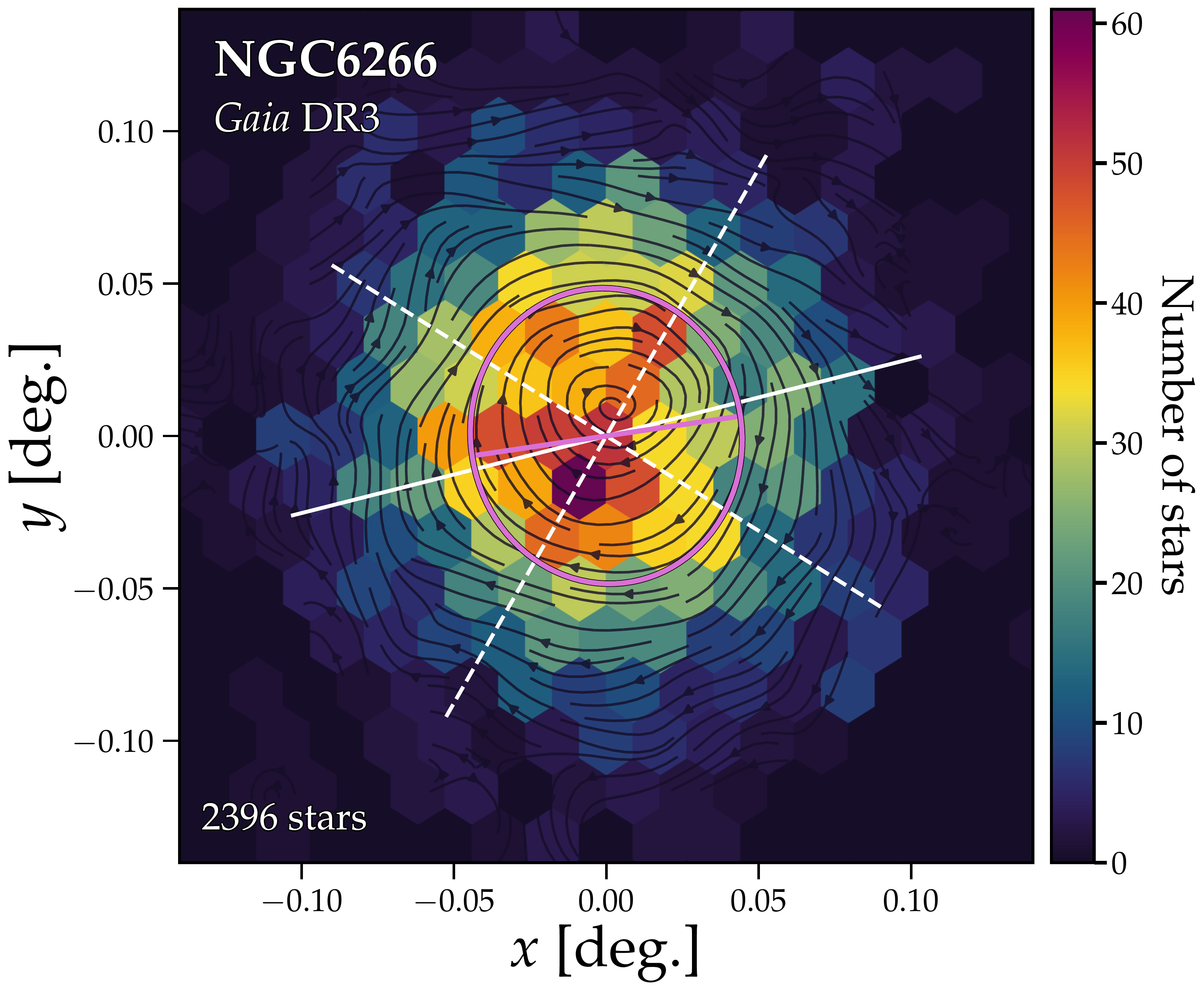}
\includegraphics[trim = 0mm 0mm 0mm 0mm, clip, width=0.33\textwidth]{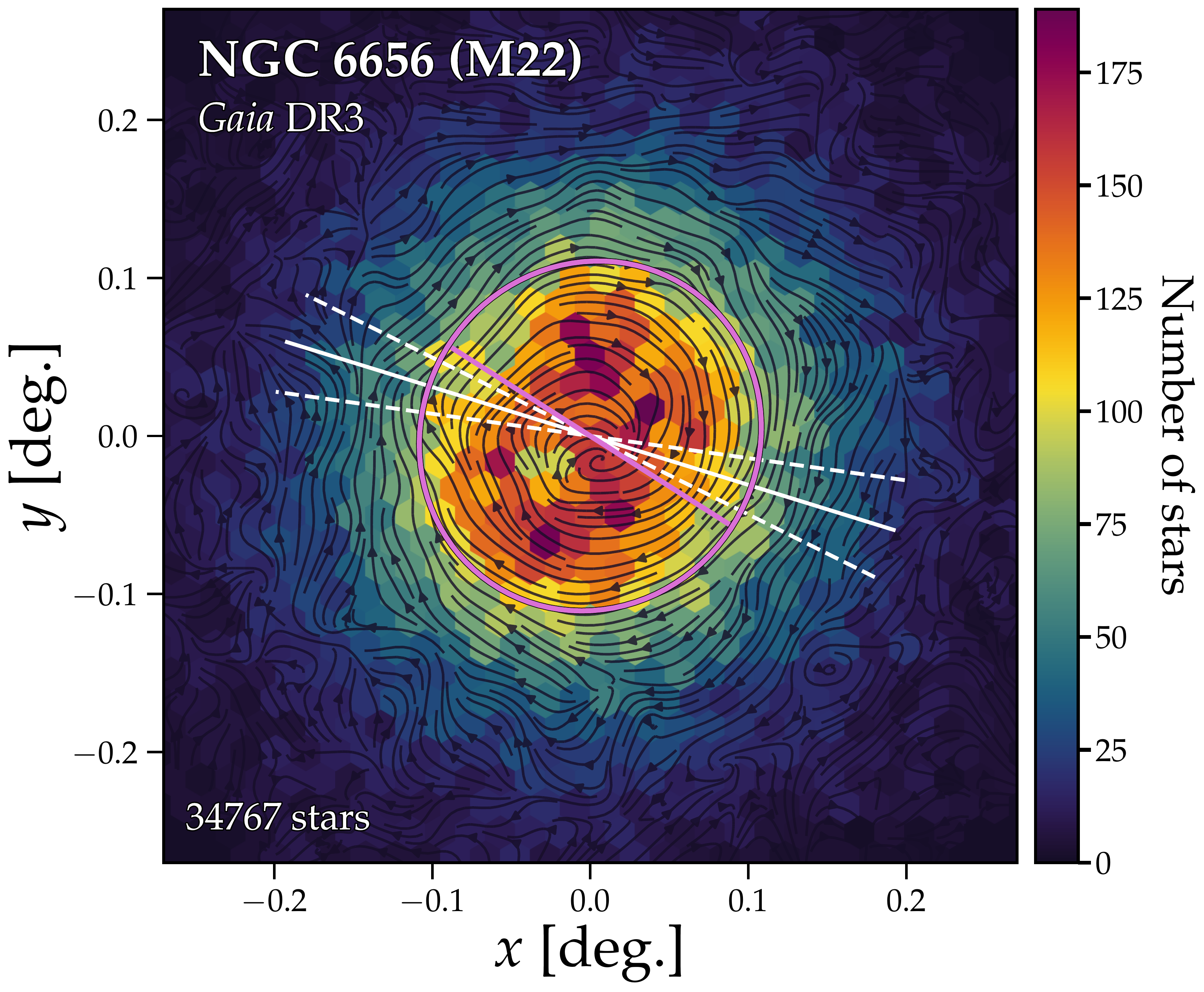}
\includegraphics[trim = 0mm 0mm 0mm 0mm, clip, width=0.33\textwidth]{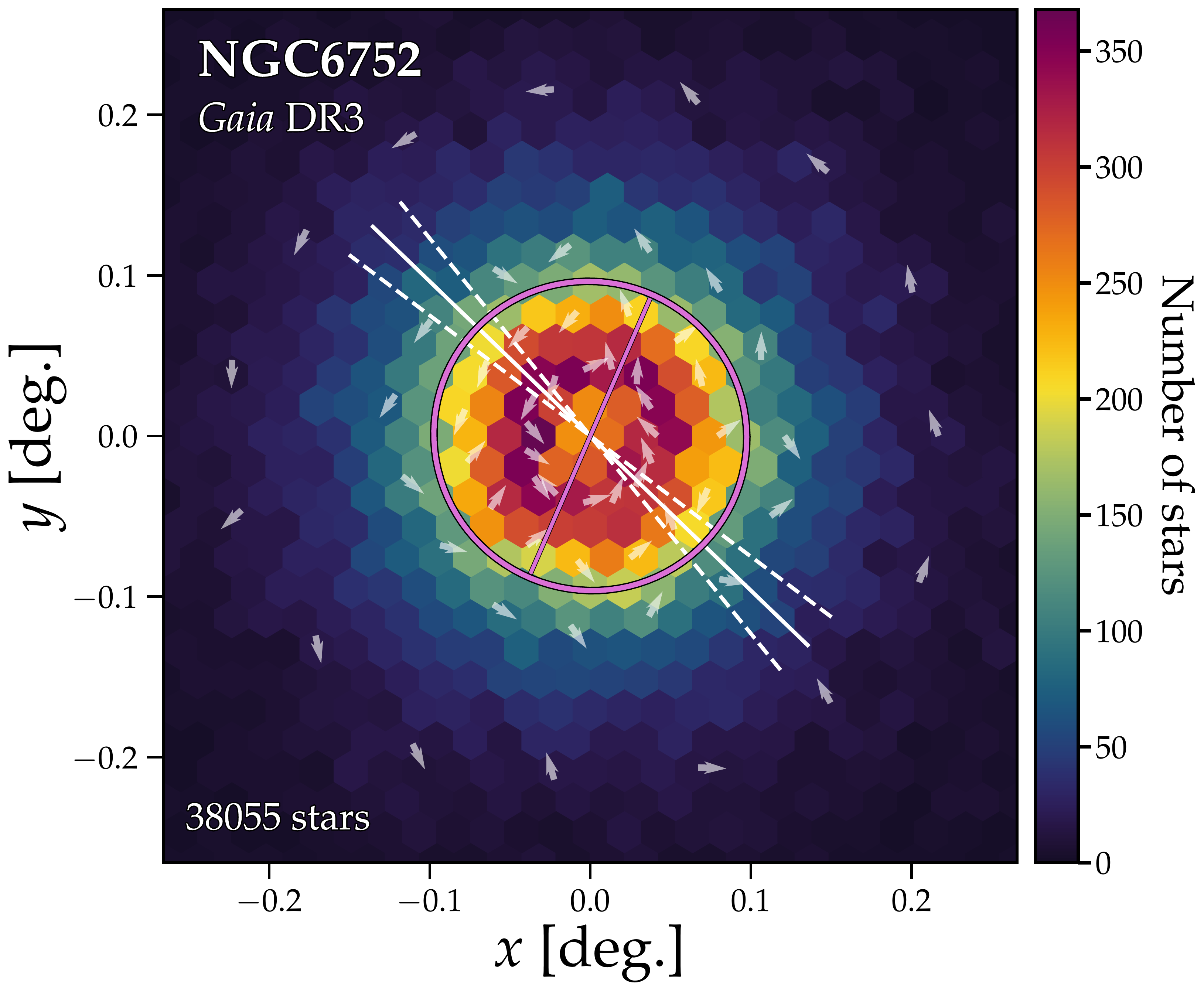}
\caption{Continuation of Fig.~\ref{fig:Streamline} for six other clusters with less pronounced rotation signals on the plane of the sky.} 
\label{fig:Streamline_cont}
\end{center}
\end{figure*} 

\section{Gamma-ray spectrum}\label{sec:gamma-ray-spectrum}

\begin{figure}
\begin{center}
\includegraphics[trim = 0mm 0mm 0mm 0mm, clip, width=0.5\textwidth]{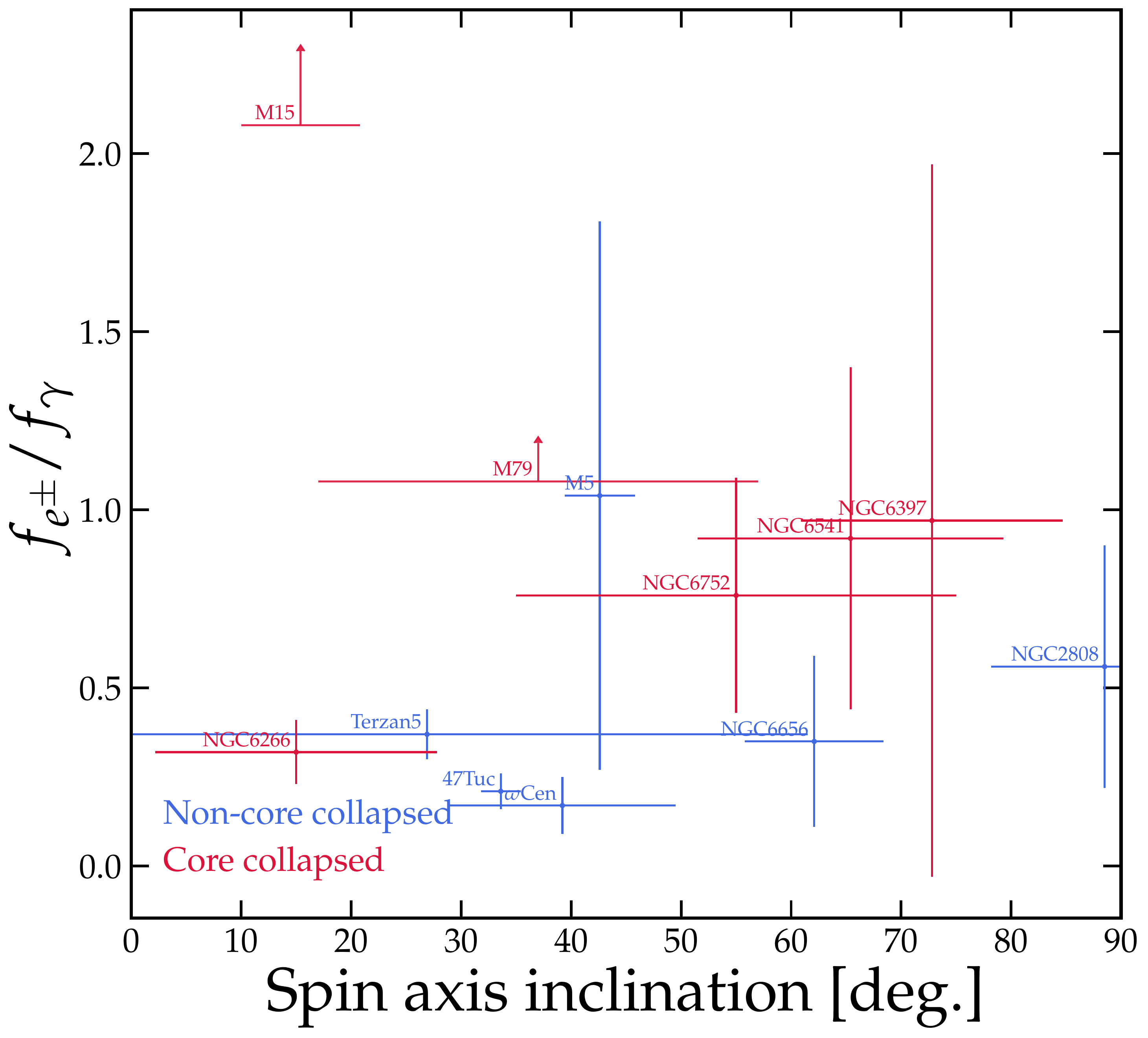}
\caption{The ratio $f_{e^\pm}/f_\gamma$ parameterises the relative contribution of inverse Compton emission relative to curvature radiation for the gamma-ray luminosity of a millisecond pulsar population. The values are obtained from a fit to the Fermi gamma-ray spectrum of the GCs in Song et al.~\cite{Song_Fermi_GCs}. The clusters with only lower limits are two cases where only a spectral fit to a falling inverse Compton emission spectrum was found.} 
\label{fig:fe_fgamma}
\end{center}
\end{figure} 

In Fig.~\ref{fig:fe_fgamma} we include an additional plot showing the relation between the ratio $f_{e^\pm}/f_\gamma$ versus the inclination angle. This ratio parameterises the relative importance of inverse Compton emission from $e^\pm$ pairs compared to curvature radiation in the gamma-ray spectrum of the population of millisecond pulsars in the GCs. If the reason why the gamma-ray luminosity from GCs decreases for larger inclination angles is because of an aligned population of millisecond pulsars, then this would imply we would see a larger contribution from curvature radiation for small inclination angles, i.e. a positive correlation between $f_{e^\pm}/f_\gamma$ and inclination---the reason being that the curvature radiation would be less isotropised compared to the inverse Compton emission. At present it is difficult to extract a clear trend due to the large uncertainty on this parameter. It is possible a very mild positive correlation could be present however the two clusters for which only \textit{lower} bounds on $f_{e^\pm}/f_\gamma$ were obtained in~\cite{Song_Fermi_GCs} were not present. However, it should be noted the quantity $f_{e^\pm}/f_\gamma$ was obtained by fitting the spectra to a sum of two spectral components over only a handful of bins in energy. From inspection of the two clusters with only lower limits published, M15/NGC7078 and M79/NGC1904~\cite[see Fig.~C1 of Song et al.,][]{Song_Fermi_GCs}, it does not appear conclusive that a purely inverse Compton spectrum is a better fit, and clearly more data and in-depth analysis of the spectrum is needed. At this stage, the evidence based on the spectral properties of the gamma-ray emission is inconclusive but could constitute an interesting test of the working hypothesis in the future.

\textbf{\textit{Acknowledgements}}.---This work was supported by The University of Sydney. CAJO is supported by the Australian Research Council under the grant number DE220100225. 
RMC acknowledges support from Australian Research Council Discovery Grant DP220102506.
AKM acknowledges the support from the Portuguese Funda\c c\~ao para a Ci\^encia e a Tecnologia (FCT) grants UIDB/FIS/00099/2020, EXPL/FIS-AST/1368/2021, and from the Caltech Division of Physics, Mathematics and Astronomy for hosting research leaves during 2017-2018 and 2019 when some of the ideas underlying this work were initially developed.
The authors thank
Arash Bahramian,
Lilia Ferrario, 
Mark Krumholz, and 
Maddie McKenzie for useful discussions
and they particularly acknowledge Simon Johnston and Holger Baumgardt for extended discussions and helpful feedback on a draft of this paper.
We also gratefully acknowledge Holger Baumgardt's database of Galactic Globular clusters: \url{https://people.smp.uq.edu.au/HolgerBaumgardt/globular/}.
This work has made use of data from the European Space Agency (ESA) mission {\it Gaia} (\url{https://www.cosmos.esa.int/gaia}), processed by the {\it Gaia} Data Processing and Analysis Consortium (DPAC,
\url{https://www.cosmos.esa.int/web/gaia/dpac/consortium}). Funding for the DPAC has been provided by national institutions, in particular the institutions participating in the {\it Gaia} Multilateral Agreement. Some of the authors are members of the {\it Gaia} DPAC.

\bibliographystyle{bibi.bst}

\bibliography{astro.bib}

\end{document}